%% file: main.tex
\documentclass[]{youtu} % For LaTeX2e

\usepackage{mathpazo}
\usepackage{graphicx}
\usepackage{natbib}
\setcitestyle{square,numbers,comma}

\usepackage{amsmath}
\usepackage{amssymb}
\usepackage{amsfonts}
\usepackage{bm}
\usepackage{booktabs}
\usepackage{multirow}
\usepackage{float}
\usepackage{placeins}
\usepackage{url}

\input{math_commands.tex}

\graphicspath{{figures/}}

\hypersetup{
    pdftitle={OmniEVIE: Evidence-Vector-Informed Embedding for Multi-modal Retrieval},
    pdfauthor={Zifei Wang, Wei Wen, Qiang Ji, Qian-Wen Zhang, Ruizhi Qiao, He Liu, Xing Sun, Qiang Lin},
    pdfsubject={Multimodal Information Retrieval},
    pdfkeywords={OmniEVIE, Multimodal Retrieval, Late Interaction, ST-HAC, Prefix-MRL}
}

\title{OmniEVIE: Evidence-Vector-Informed Embedding for Multi-modal Retrieval}

\author{\small Zifei Wang$^{1,*}$, Wei Wen$^{2,*\dagger}$, Qiang Ji$^{1}$, Qian-Wen Zhang$^{1,2}$, Ruizhi Qiao$^{2}$, He Liu$^{1}$, Xing Sun$^{2}$, Qiang Lin$^{2}$}
\affiliation{$^{1}$Tencent IMA Product Center \quad $^{2}$Tencent Youtu Lab}

\date{November 19, 2026}
\correspondence{jawnrwen@tencent.com}
\sourcecode{https://github.com/Tencent/OmniEVIE}
\model{https://huggingface.co/tencent/OmniEVIE-4B}
\youtufinalcopy

\begin{document}

\abstract{Multimodal retrieval now has to match dense pages, fine-grained entities, ordered video, and the tools and memory an agent looks up. A single vector drops the local evidence that decides these matches. A cross-encoder can recover it, but only by rereading every recalled page or clip. Late interaction keeps the evidence in the index; storing every token vector is too large to serve. OmniEVIE is a 4.5B multi-vector embedding model that covers text, images, visual documents, video, and tools with one compressed index and a three-stage retriever. Each record holds at most $128$ int8 vectors of dimension $64$: $9{,}472$ bytes per record, or $8.82$ GiB per million records. Retrieval runs on that compressed index, and a returned hit names the page region or the video interval. Training, merging, indexing, and retrieval are designed together. Every task is written as one retrieval sample. A three-grade evidence check removes false negatives, and the sampling rate follows the gap between the baseline and the teachers. Candidates are compressed with ST-HAC during training, in the form stored in the index, so the training score and the serving score use the same vectors. Stage-wise runs and specialist branches are merged by parameter block. A saliency gate and a noise floor rewrite the query score. One compressed record supplies a mean vector, $1$-bit signs, and int8 multi-vectors, used for the coarse, middle, and final stages. Trained on public datasets, the full sequence at $d=2048$ scores $60.68$ on MMEB-V3 (190 tasks), and the $K128 \times d64$ int8 index scores $60.06$. The same compressed index scores $77.96$ on MMEB-V2 (78 tasks) and $62.96$ on ViDoRe V3 (48 tasks). On corpora of $1$ million and $5$ million records, one single-thread query takes $54.2$ ms and $120.2$ ms, $11.1\times$ and $25.3\times$ faster than exact search over $4096$-dimensional single vectors, and $236\times$ and $532\times$ faster than full int8 MaxSim. Video is encoded directly from frames and a hit is read back to its start and end in seconds, so neither indexing nor localization uses a video parsing model. On the $1{,}083$ queries of the QVHighlights validation set, the heat peak falls inside the labeled segment for $78.0\%$ of them and the returned span intersects it for $94.3\%$, against $44.9\%$ for a random span of the same length; the span covers $18\%$ of the video on average. Weights and code are at \url{https://huggingface.co/tencent/OmniEVIE-4B} and \url{https://github.com/Tencent/OmniEVIE}.}

\maketitle

\renewcommand{\thefootnote}{\fnsymbol{footnote}}
\footnotetext[1]{Equal contribution.}
\footnotetext[2]{Corresponding author. Contact: \texttt{jawnrwen@tencent.com}.}
\renewcommand{\thefootnote}{\arabic{footnote}}
\setcounter{footnote}{0}

\section{Introduction}

\subsection{From a single vector to multi-vector retrieval}

Retrieval queries now include multi-page reports, fine-grained products and landmarks, long videos with ordered actions, and agent contexts built from tool descriptions, interface state, and multi-step history \citep{jiang2024vlm2vec,meng2025vlm2vec,huang2026mmebv3}. Relevance often depends on a few non-adjacent pieces of evidence: a header and one cell in a financial table, the order of two actions in a video, or one parameter in a long tool document.

A single $d$-dimensional vector scores a query and a candidate by one inner product. \citet{weller2026theoretical} show that this score can place only some combinations first: past a corpus size that grows polynomially with $d$, a document that must be recognized by several pieces at once cannot be ranked first by any query. Adding data or enlarging the backbone does not remove that limit.

Late interaction keeps a query token set $Q=\{q_1,\ldots,q_{L_q}\}$ and a candidate set $D=\{v_1,\ldots,v_K\}$, and scores each query token by its best match in the candidate \citep{khattab2020colbert,faysse2025colpali}:
\begin{equation}
\operatorname{MaxSim}(Q, D) = \sum_{i=1}^{L_q} w_i \max_{1 \le j \le K} \langle q_i, v_j \rangle,
\label{eq:maxsim}
\end{equation}
with $w_i>0$. Because each query token can land on a different span, the score is no longer limited by the dimension of one vector. The leading visual-document and text-to-video retrievers use this form \citep{faysse2025colpali,moreira2026nemotron,reddy2025videocolbert,wan2025clamr}.

\subsection{Three obstacles at serving time}

That representation is expensive to serve, for three separate reasons. A high-resolution page or a clip of tens of seconds produces thousands of vectors, so a million records occupy hundreds of gigabytes in floating point. A query then has to be compared with every vector: one single-thread query on a million-record corpus takes on the order of ten seconds. Failure modes also differ across documents, images, video, and tools: upweighting one modality often degrades the others, and a model trained on the full token sequence loses accuracy once those tokens are compressed for the index.

\subsection{What OmniEVIE does}

The usual compromise retrieves with one vector and reranks with a cross-encoder \citep{moreira2026nemotron,li2026qwen3vl}. Each of the $k$ recalled pages or clips is forwarded again with the query, so the system has to read the raw media online. An uncompressed multi-vector index has the opposite problem: prior studies show that MaxSim selects only about $1\%$ of the stored vectors \citep{qin2026compression}. OmniEVIE keeps the width at $d \le 128$, bounds the number of vectors, and stores them at low bit width. Search runs on that compressed index (Table~\ref{tab:deploy}).

The model extends the evidence check and the local multi-vector index of EVIE from visual documents to general multimodal retrieval \citep{wang2026evie}. A mined candidate stays a negative only when it lacks the evidence the query asks for. Position codes guide clustering and are then discarded, so the index stores semantic centroids. The same integer record is what the retriever filters and what it uses to name the hit. This release is trained on $57$ public datasets. The changes are in the backbone, the order in which modalities are added, and the decision to train on the same clustered vectors that are written to the index.

\begin{figure}[t]
\centering
\includegraphics[width=\linewidth]{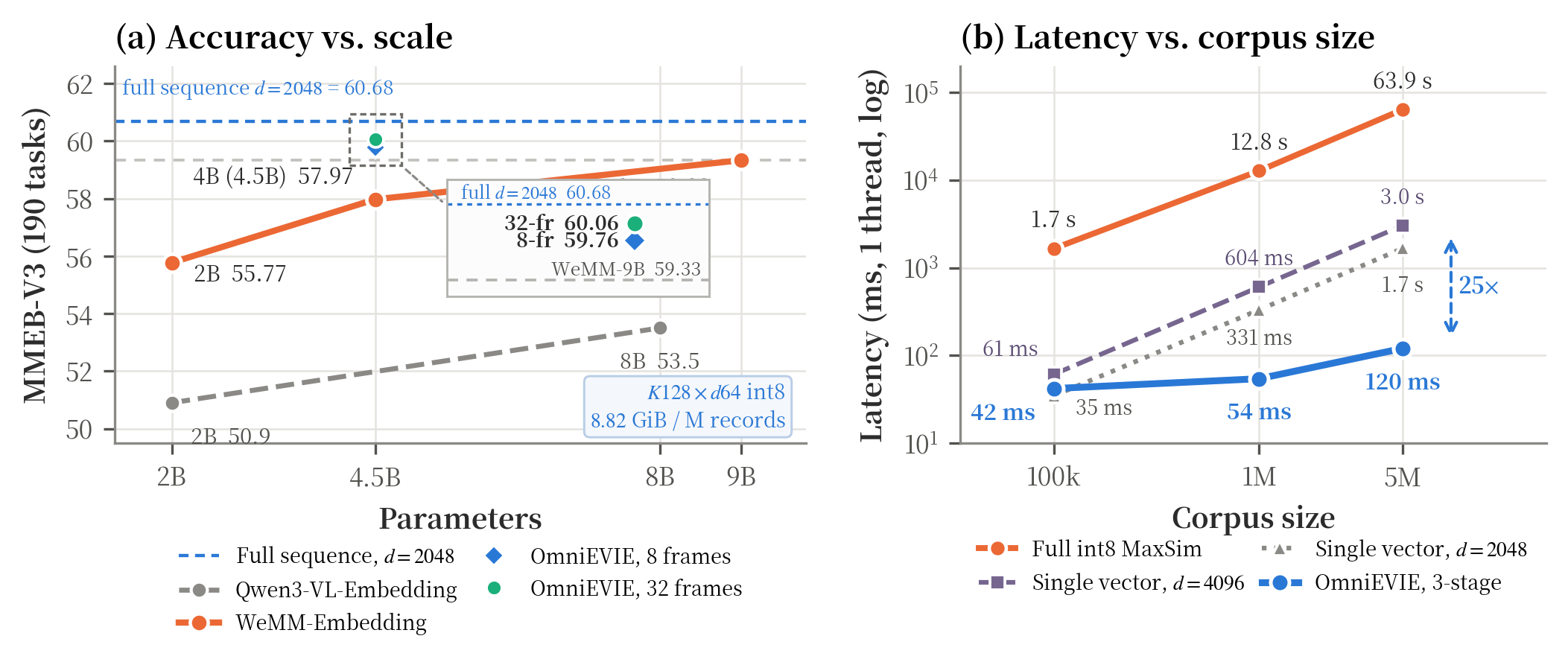}
\caption{Accuracy and latency of OmniEVIE. (a) MMEB-V3 mean over 190 tasks against parameter scale. The blue line is the full sequence at $d=2048$: the same weights keep every token vector, with no ST-HAC and no int8 quantization, and score $60.68$. The circle and the diamond are the released $K128 \times d64$ int8 index, $8.82$ GiB per million records, at $60.06$ for $32$ frames and $59.76$ for $8$ frames. WeMM-Embedding-4B and OmniEVIE share a $4.5$B backbone. (b) Single-thread latency against corpus size for full int8 MaxSim, exact single-vector search at $d=2048$ and $d=4096$, and the three-stage retriever, whose first stage is a $d=64$ mean vector.}
\label{fig:overview}
\end{figure}

A hit also carries the place of its evidence. A page or an image returns the bounding rectangle of the hit clusters, and a video returns the start and end in seconds of the hit centroids, so the downstream generator reads that region or that span instead of the whole page or video. The video side encodes the sampled frames directly and takes time from the frame indices, so neither indexing nor localization goes through a parsing model that transcribes video into text. The localization numbers are in Section~\ref{sec:analysis-loc}. Figure~\ref{fig:overview} shows the resulting accuracy and latency, and Table~\ref{tab:deploy} compares the four serving designs.

\begin{table}[t]
\centering
\scriptsize
\setlength{\tabcolsep}{3pt}
\caption{Serving comparison of four multimodal retrieval designs.}
\label{tab:deploy}
\begin{tabular}{p{0.16\linewidth}p{0.19\linewidth}p{0.22\linewidth}p{0.18\linewidth}p{0.18\linewidth}}
\toprule
 & Single vector & Single vector + reranker & Full multi-vector & OmniEVIE \\
\midrule
Score & One inner product & Coarse score, then cross-attention & Full-token MaxSim & Fixed multi-vector, \mbox{$K\le 128$}, $d=64$ \\
Local hit & None & No coordinates & Per-token hit & Page rectangle and video seconds; span hit $94.3\%$ \\
Million-scale index & A few GiB of vectors & Must read raw media for reranking & Hundreds of GiB & $7.87$--$8.82$ GiB \\
Online forwards & Encode the query once & One query encode plus $k$ media forwards & Encode the query once & Encode the query once \\
Bottleneck & Dimension of one vector & Rerank tail latency & Index size and full scan & Integer and bit operations \\
Width & One trained width, or a separate MRL head & Reranker maintained apart from the retriever & Fixed high dimension & One Prefix-MRL head, six widths, ST-HAC at training time \\
\bottomrule
\end{tabular}
\end{table}

The rest of the report follows that record from training to search. Section~\ref{sec:data} writes retrieval, classification, question answering, grounding, and tool use as one sample, grades the evidence in mined negatives, and adds video negatives that differ only in order. Section~\ref{sec:model} trains a bidirectional backbone with native video tokens and one Prefix-MRL head for widths $64$ through $2048$; clustering during training and the query-side score are in that section. Section~\ref{sec:merge} adds stage-wise and specialist checkpoints by parameter block. Section~\ref{sec:sthac} uses one spatiotemporal clustering for every modality; the bytes in one record are given there. Section~\ref{sec:retrieve} derives three representations from one record and narrows the candidate window. Section~\ref{sec:loc} reads the hit back to a page rectangle or to seconds in a video without a video parsing model. The numbers are in Section~\ref{sec:analysis-loc}.

\section{Related work}

\subsection{General multimodal embeddings and multi-vector representations}

One general embedding model is now expected to handle text, images, documents, and video. VLM2Vec and the MMEB suites put retrieval, classification, question answering, and grounding into one evaluation \citep{jiang2024vlm2vec,meng2025vlm2vec,huang2026mmebv3}. \citet{esfandiarpoor2026agentic} measure what that similarity score costs once the query needs several pieces of evidence. Qwen3-VL-Embedding, WeMM-Embedding, Ovis-Embedding, and U-MARVEL train a single vector with multi-stage contrastive learning, hard negatives, reranker supervision, and cross-scale distillation \citep{li2026qwen3vl,zhou2026wemm,ovis2026embedding,li2026umarvel}. By contrast, jina-embeddings-v4 and v5-omni keep both a single vector and a multi-vector output, and lock the vision tower so the geometry stays aligned \citep{gunther2025jina4,hoenicke2026jina5}.

Late interaction was established for text by ColBERT and ColBERTv2 \citep{khattab2020colbert,santhanam2022colbertv2} and carried to full-page visual documents by ColPali, which replaces an OCR pipeline with a page screenshot \citep{faysse2025colpali}. ViDoRe V3 extends that evaluation to multilingual pages \citep{loison2026vidore}. Jina-ColBERT-v2 covers multilingual text \citep{jha2024jinacolbert}. ModernVBERT and Nemotron-ColEmbed-V2 study bidirectional attention and model merging for document late interaction \citep{teiletche2025modernvbert,moreira2026nemotron}. Video-ColBERT and CLAMR extend late interaction to video \citep{reddy2025videocolbert,wan2025clamr}. EVIE applies evidence checking, a Prefix-MRL head, and layout-aware HAC to visual documents \citep{wang2026evie}. Matryoshka representation learning and MatFormer nest several widths in one forward pass \citep{kusupati2022mrl,devvrit2023matformer}. MetaEmbed changes the number of meta-tokens at test time \citep{xiao2026metaembed}. OmniEVIE carries this design to images, video, text, and agent tasks, and adds a shared spatiotemporal compression, block-wise merging, query-side scoring, and million-scale three-stage retrieval.

\subsection{Compression and fast multi-vector search}

Compressing a multi-vector index is a choice of which part of the match to keep. MUVERA replaces MaxSim by one fixed-dimensional vector so that an existing maximum-inner-product index can be reused \citep{dhulipala2024muvera}. PLAID keeps the multi-vector score but filters through a global codebook and residual codes \citep{santhanam2022plaid}, and query-aware token budgeting delays the choice of raw tokens until the final stage \citep{rishi2026tokenbudget}. A third line reduces the number of stored vectors, either by merging tokens inside the encoder \citep{bolya2023tome} or by clustering and pruning a frozen index, as in attention-guided clustering \citep{qin2026compression}, MarginMerge \citep{mahdizadeh2026marginmerge}, and prune-then-merge \citep{yan2026sculpting}. ST-HAC belongs to this third line, with one difference: the same clustering is used while training and again when the index is written. The mean, the $1$-bit signs, and the int8 multi-vectors are then three readings of that one record, at a fixed int8 scale.

\subsection{False negatives, rank distillation, and model merging}

The quality of a contrastive embedding depends on which negatives are false and on whose ranking is distilled. FALCON studies false negatives in large-scale image-text pretraining \citep{kim2025falcon}. CORE distills a reranker ranking into an embedding with Rank-KL \citep{song2026core}. On the parameter side, task arithmetic adds and subtracts task vectors \citep{ilharco2023taskarith}, and TIES-Merging drops small deltas and resolves sign conflicts \citep{yadav2023ties}. Conan-embedding-v3, PlaM, and training-free mitigation study modality-specific fusion, layer-selective injection, and the recovery of language ability after multimodal tuning \citep{li2026conan,wang2026plam,ratzlaff2024mitigation}. The merge in OmniEVIE gives each parameter block its own coefficient, so the merger and the projection head can take a different step from the language layers.

\section{Training data}
\label{sec:data}

OmniEVIE puts every task into one retrieval sample. The instruction says which task it is, the evidence grade says whether a mined negative is truly negative, and a dynamic sampling rule adjusts how often each sample is revisited.

\subsection{One sample format}

Each example is
\begin{equation}
z_i = \big(I_i,\, q_i,\, d_i^{+},\, \mathcal{N}_i,\, y_i\big),
\label{eq:sample}
\end{equation}
where $I_i$ is a scene instruction, $q_i$ is text, an image, an image-text pair, or a video, $d_i^{+}$ is the positive, $\mathcal{N}_i$ is a set of explicit hard negatives, and $y_i$ is a teacher score or a distribution. Retrieval and grounding use the page, image, crop, or clip as the positive. Classification uses the class name as a candidate and prefers near-miss classes from the same dataset, balanced by class frequency. Open question answering pairs the question and the image with the reference answer. Multiple choice uses the correct option as the positive and the other options as hard negatives. Tool retrieval uses the tool document or skill description as the candidate. Interface and memory tasks are held out for zero-shot evaluation.

\begin{figure}[t]
\centering
\includegraphics[width=\linewidth]{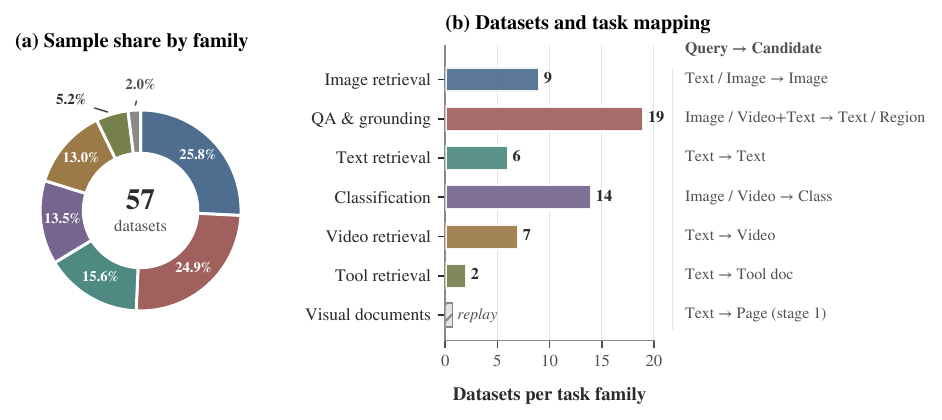}
\caption{Training mixture across $57$ public datasets (about $2$M samples). (a) Sample share by task family: image retrieval $25.8\%$, QA and grounding $24.9\%$, text retrieval $15.6\%$, classification $13.5\%$, video retrieval $13.0\%$, tool retrieval $5.2\%$, and visual documents $2.0\%$. (b) Dataset count and query-to-candidate mapping per family; visual documents are replayed from stage 1.}
\label{fig:data}
\end{figure}

A batch contains one task, so every in-batch negative is another candidate of that task. The training mixture comprises about $2$M samples from $57$ public datasets (Figure~\ref{fig:data}).

\subsection{Scene instructions}

Image-to-image search takes the instruction, the image, and an empty text span as the query. Composed search places the real question after the image. A text query places the instruction in front of the question, and the remaining zero-shot tasks share one task prefix (Table~\ref{tab:instruct}). Very thin or very small images are resized on the short side and padded so that the aspect ratio stays bounded.

\begin{table}[t]
\centering
\scriptsize
\setlength{\tabcolsep}{3pt}
\caption{Scene instructions and query layout.}
\label{tab:instruct}
\begin{tabular}{p{0.22\linewidth}p{0.24\linewidth}p{0.32\linewidth}p{0.16\linewidth}}
\toprule
Scene & Query & Instruction & Candidate \\
\midrule
Image to image & Instruction + image, empty text & Find an image of the same object or entity & Image \\
Composed image-text & Instruction + image + question & Find an image that answers the question about the given image & Image \\
Text to image or page & Instruction + query text & Retrieve the image or page the request names & Image or page \\
Text to video & Instruction + query text & Retrieve the video the request names; the instruction is prepended & Native video \\
Other zero-shot tasks & Task instruction + query & One shared task prefix & Text, image, or video \\
\bottomrule
\end{tabular}
\end{table}

\subsection{Evidence grades and temporal hard negatives}

High-scoring mined candidates are the usual source of hard negatives, and they are also where missed positives hide. If a page that answers the query is treated as a negative, the gradient pushes that page away from the query. A multimodal judge assigns each mined negative one of three grades (Table~\ref{tab:evidence}). Candidates with full evidence are promoted to positives (each receiving weight $\omega_i = 1/n_i^{+}$ when a query has $n_i^{+}$ positives), those with partial evidence are masked in the loss denominator, and those with no evidence remain hard negatives.

\begin{table}[t]
\centering
\small
\setlength{\tabcolsep}{3pt}
\caption{Three-grade evidence check on the visual-document pool. Counts are candidate negatives before the grade is applied.}
\label{tab:evidence}
\begin{tabular}{p{0.22\linewidth}p{0.28\linewidth}p{0.42\linewidth}}
\toprule
Grade & Meaning & Treatment \\
\midrule
Full evidence & The candidate answers the query & Promoted to a positive. Of $775{,}635$ queries, $175{,}137$ gain an extra positive. \\
Partial evidence & A useful clue, not a full answer & Masked in the denominator. Together with full evidence, $223{,}877/1{,}551{,}270$ ($14.4\%$) of candidate negatives are masked. \\
No evidence & Irrelevant or contradictory & Kept. $1{,}327{,}393$ hard negatives remain. \\
\bottomrule
\end{tabular}
\end{table}

Static background is often enough to recognize a clip, so two hard negatives put the loss on temporal order. A reversed negative pairs a forward clip (positive) with the same clip in reverse, and the query pairs a forward description with a reversed one. An off-window negative keeps the target span as the positive and takes another span of the same long video as the negative, under the original query.

\subsection{Dynamic sampling from a baseline and several teachers}

Not every sample deserves the same update. Some are already solved by the baseline, and on some tasks every available teacher is weak. The teacher is chosen by the form of the answer. Open answers and fine-grained classification use a multimodal embedding teacher and its in-batch similarity distribution. Ranking tasks, including text-to-image, video grounding, and tool retrieval, use a cross-encoder and the gaps among candidates. Multiple choice uses a discriminative teacher and the distribution over options.

Each sample is then labeled by whether the baseline and the teacher place the positive first (Table~\ref{tab:five}). Samples that both models rank correctly by a wide margin are downsampled. Samples the teacher ranks correctly and the baseline misses are upweighted. When the baseline beats the teacher, distillation is turned off and the loss uses the gold label. If the baseline outperforms the teacher on more samples of a task than the reverse, distillation is turned off for that task.

\begin{table}[t]
\centering
\small
\setlength{\tabcolsep}{3pt}
\caption{Five-way sample policy. ``First'' means the positive is ranked first. Distillation is on only when the teacher is correct.}
\label{tab:five}
\begin{tabular}{p{0.18\linewidth}p{0.16\linewidth}p{0.11\linewidth}p{0.26\linewidth}p{0.20\linewidth}}
\toprule
Class & Baseline & Teacher & Keep rule & Hard negatives \\
\midrule
1, easy agreement & First, large margin & Correct & Fixed downsample; class tasks keep a per-class floor & Baseline picks plus the original set \\
1b, borderline & First, small margin & Correct & Keep all & Same as class 1 \\
2, teacher fills a gap & Not first & Correct & Upweight & Same as class 1 \\
3, baseline ahead & First & Wrong & Keep all, gold label only & Original set only \\
4, hard & Not first & Wrong & Keep all, gold label only & Original set only \\
\bottomrule
\end{tabular}
\end{table}

\section{Model and training}
\label{sec:model}

\subsection{Backbone, bidirectional attention, and native video}

OmniEVIE is built on the Qwen3.5-VL mixed-attention backbone: a vision encoder, a cross-modal merger, language layers, and a Prefix-MRL head. The released model has $4.5$B parameters and hidden width $d_{\mathrm{hid}}=2560$. Language layers alternate GatedDeltaNet blocks and full attention. The document stage distills an $8$B bidirectional EVIE teacher with $d_{\mathrm{hid}}=4096$ into this $4.5$B model \citep{wang2026evie}.

Text is embedded by the tokenizer. Images and pages become patch tokens after the vision encoder and a spatial merge. A causal mask would hide a later instruction from the visual tokens that precede it. Full-attention layers use bidirectional attention, and the mask marks padding. Linear-attention blocks stay as in the backbone. Every token can then attend over the full sequence \citep{moreira2026nemotron,wang2026evie,teiletche2025modernvbert}.

OmniEVIE feeds each clip as one sequence, so adjacent frames meet inside the encoder. Per-frame encoding instead runs one forward per frame and pools the vectors afterward, so frames meet only after the backbone, a reversed clip yields the same set of frames, and exceeding the token budget forces dropping frames. Adjacent frames are packed at the vision-encoder input with temporal stride $\Delta t_{\mathrm{patch}}=2$, and the last frame is repeated when the count is odd. Each temporal group is prefixed by a timestamp in seconds. When the token budget is exceeded, every frame is scaled down by the same factor and the time samples stay. Timestamps increase with position, so a reversed clip pairs content with a different timestamp, and the reversed negatives of Section~\ref{sec:data} have a nonzero gradient.

\subsection{Prefix-MRL}

Distillation needs a wide vector, and the index needs a narrow one. One shared matrix $W \in \mathbb{R}^{2048 \times 2560}$ sits on top of the backbone. For a final hidden state $h \in \mathbb{R}^{2560}$ and a width $d \in \mathcal{D}=\{64,128,256,512,1024,2048\}$,
\begin{equation}
e^{(d)}(h) = \operatorname{L2}\!\left(W_{:d}\, h\right) = \frac{W_{:d}\, h}{\|W_{:d}\, h\|_{2}} \in \mathbb{S}^{d-1}.
\label{eq:mrl}
\end{equation}
A single forward pass produces every width, and a low-width loss updates only the first $d$ rows. Serving always stores $d=64$, while the $d=2048$ branch provides the high-width reference for self-distillation.

\subsection{Four training stages}

\begin{figure}[t]
\centering
\includegraphics[width=\linewidth]{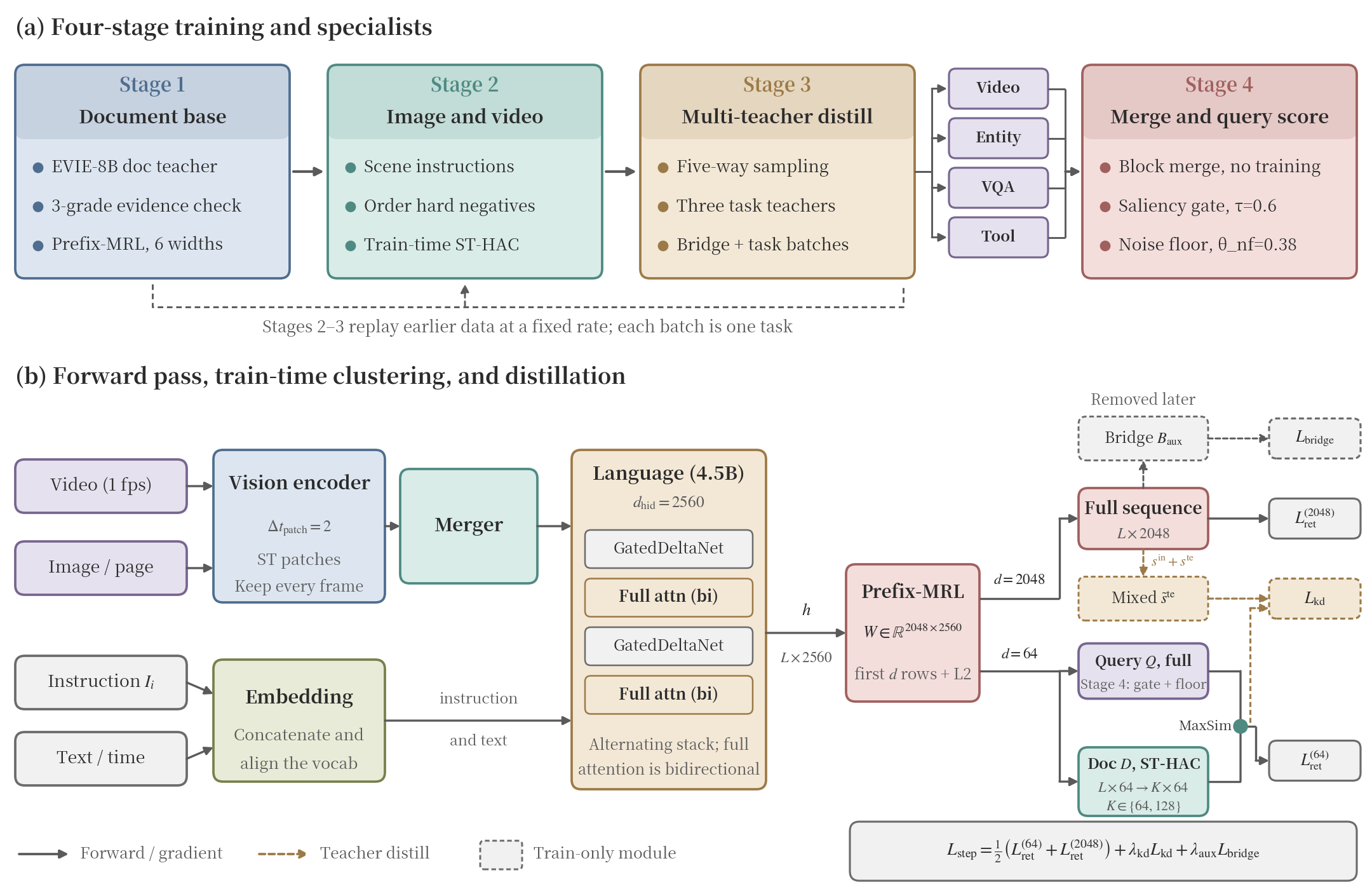}
\caption{Four-stage schedule and the forward graph. (a) The curriculum and the four specialist branches split off after stage 3. (b) Vision, instruction, and text pass through the backbone and the Prefix-MRL head, then split into a $d=2048$ reference branch and a $d=64$ serving branch. The query keeps its full sequence, and stage 4 attaches the gate and the noise floor. Candidates are clustered with ST-HAC during training. The $d=2048$ branch and an external teacher distill the $d=64$ branch.}
\label{fig:train}
\end{figure}

Training every modality from scratch in one stage lets frequent easy tasks crowd out the harder alignment, and adding a modality later can erase the document behavior already learned. OmniEVIE adds modalities in four stages (Figure~\ref{fig:train}, Table~\ref{tab:stages}) and replays earlier data at a fixed rate. Each batch is still a single task.

\begin{table}[t]
\centering
\scriptsize
\setlength{\tabcolsep}{3pt}
\caption{The four training stages and what each stage replays.}
\label{tab:stages}
\begin{tabular}{p{0.18\linewidth}p{0.24\linewidth}p{0.30\linewidth}p{0.22\linewidth}}
\toprule
Stage & Input & Supervision & Replay \\
\midrule
1, documents & Bidirectional document vectors and the six Prefix-MRL widths & Listwise distillation from EVIE-8B, the three-grade check, joint training of the six widths & Document pool for later stages \\
2, image and video & Image-to-image, composed image-text, text-to-image, native video & Scene instructions, temporal hard negatives, a one-way ranking constraint, ST-HAC on the candidates in the batch & Document and earlier data at a fixed rate; one batch is one task \\
3, all tasks & Classification, question answering, grounding, long text, tools, interfaces, memory & The five-way sampler, three teachers by task, a projection bridge to the high-width teacher & Replay quotas kept; batches stay single-task \\
4, merge and query score & Specialists added back; the query-side score is trained & Block-wise merge and the gate with the noise floor & Deltas added on the blocks they were trained on \\
\bottomrule
\end{tabular}
\end{table}

\subsection{Loss, distillation, and clustering during training}

Both the $d=64$ branch and the $d=2048$ branch are trained. Candidates on the narrow branch are clustered as they will be at indexing time, and the score uses those centroids.

The candidate pool of a query is the other documents in the batch plus its own hard negatives. Three kinds of candidate are excluded from the negative set (other positives of the same query stay positives, samples from another source leave this loss, and candidates scored at least as high as the positive by the teacher leave the negatives), and every remaining candidate in the pool is a negative. Let $\pi_i$ be the softmax probability of the positive. A focal weight emphasizes queries that are not yet separated.

\begin{equation}
\mathcal{L}_{\mathrm{focal}} = \mathbb{E}_{\mathrm{batch}}\big[a_i\,(-\log \pi_i)\big],
\quad
a_i = \frac{\operatorname{sg}\!\big((1-\pi_i)^{\gamma}\big)}{\mathbb{E}_{\mathrm{batch}}\!\big[\operatorname{sg}((1-\pi_j)^{\gamma})\big]},
\quad \gamma = 1.
\label{eq:focal}
\end{equation}
$\operatorname{sg}$ stops the gradient through $a_i$, so the weight is a constant with respect to the parameters. Dividing by the batch mean keeps the average weight at $1$, so the scale of the loss does not follow how hard the batch is. Bidirectional samples add the reverse cross-entropy $\mathcal{L}_{d \to q}$ with weight $\beta=0.5$. Ordered pairs, including a forward clip against its reverse and negatives that have been graded, add a pairwise term $\mathcal{L}_{\mathrm{ord}}$. The gradient through the higher score $s_0$ is stopped, so only the lower score $s_1$ is decreased. The retrieval loss on one branch is
\begin{equation}
\mathcal{L}_{\mathrm{ret}} = \mathcal{L}_{\mathrm{focal}} + \beta\,\mathcal{L}_{d \to q} + \lambda_{\mathrm{ord}}\,\mathcal{L}_{\mathrm{ord}}.
\label{eq:ret}
\end{equation}

The distillation target mixes the model's internal $d=2048$ scores $s^{\mathrm{in}}$ with external teacher scores $s^{\mathrm{te}}$. Both are standardized on the valid candidates. With mixture weight $w_e$, the target is placed back on the scale of $s^{\mathrm{in}}$:
\begin{equation*}
\tilde{s}^{\mathrm{te}} = \mu_{\mathrm{in}} + \sigma_{\mathrm{in}}\big((1-w_e)\hat{z}(s^{\mathrm{in}}) + w_e\hat{z}(s^{\mathrm{te}})\big).
\end{equation*}
The distillation loss is a KL divergence between the teacher and student rankings plus a margin term,
\begin{equation}
\mathcal{L}_{\mathrm{kd}} = c_i\Big[\operatorname{KL}\!\big(p^{\mathrm{te}} \,\|\, p^{\mathrm{stu}}\big) + \lambda_{\mathrm{margin}}\,\mathcal{L}_{\mathrm{margin}}\Big],
\label{eq:kd}
\end{equation}
where $c_i$ is larger when the teacher distribution has lower entropy.

An external teacher may have dimension $d_T=4096$, which cannot be scored directly against a $2048$-dimensional student vector. A bias-free linear map $B_{\mathrm{aux}} \in \mathbb{R}^{d_T \times 2048}$, initialized orthogonally, sends the student mean vector into the teacher space. The point term is one minus the cosine. The batch term is a smoothed $\ell_1$ loss between the query--document cosine matrices. $B_{\mathrm{aux}}$ is discarded after training, and inference uses the Prefix-MRL head.

At each step the narrow branch clusters candidates using budget $K=64$ with probability $0.3$ and $K=128$ with probability $0.7$. Cluster assignment is held fixed. The mean inside a cluster and the normalization that follows it receive gradient. Every token in a cluster receives the same share $1/|C_k|$ of the centroid gradient. The step loss is
\begin{equation}
\mathcal{L}_{\mathrm{step}} = \frac{1}{2}\big(\mathcal{L}_{\mathrm{ret}}^{(64)} + \mathcal{L}_{\mathrm{ret}}^{(2048)}\big) + \lambda_{\mathrm{kd}}\,\mathcal{L}_{\mathrm{kd}} + \lambda_{\mathrm{aux}}\,\mathcal{L}_{\mathrm{bridge}}.
\label{eq:step}
\end{equation}
The $d=64$ branch scores centroids. The $d=2048$ branch scores the full sequence.

\subsection{Query-side saliency gate and noise floor}

Uniform MaxSim, with $w_i=1$ on every non-pad token, fails on multimodal queries for two reasons. Visual patches and background tokens dominate the sum, and every query token can find a medium inner product on any candidate. Those medium matches add up and dilute the real hit.

The gate starts from the uniform weights $w_i^{(0)}$. Its input is ten domain-free features $\phi_i$ of each token: the pre-normalization norm, deviation from the query mean direction, difference from the neighboring token, uniqueness inside the query, modality and position marks, and the query length. A two-layer SiLU network produces a score $g_i$. A linear term on the token direction carries an $\ell_2$ penalty. The weights are then
\begin{equation}
m_i = \exp\!\big(\tau_{\mathrm{gate}} \tanh(g_i)\big),
\qquad
w_i = \frac{L_q\, w_i^{(0)} m_i}{\sum_{j=1}^{L_q} w_j^{(0)} m_j}.
\label{eq:gate}
\end{equation}
The gate output layer is initialized at zero, so training starts from uniform MaxSim. With $\tau_{\mathrm{gate}}=0.6$, each token is rescaled between about $0.55$ and $1.8$, and the weights still sum to $L_q$. The gate has $257$ parameters.

Let $r_i = \max_{v \in D}\langle \hat{q}_i, v \rangle$. The noise-floor score amplifies the part of $r_i$ above $\theta_{\mathrm{nf}}$:
\begin{equation}
s_{\mathrm{tok}}(q_i, D) = r_i + \beta_{\mathrm{nf}}\,\operatorname{ReLU}\!\big(r_i - \theta_{\mathrm{nf}}\big),
\qquad
s(Q, D) = \sum_{i=1}^{L_q} w_i\, s_{\mathrm{tok}}(q_i, D).
\label{eq:nf}
\end{equation}
Matches below the floor are unchanged. Matches above it are multiplied by $1+\beta_{\mathrm{nf}}$ on the excess only. $\beta_{\mathrm{nf}}=0$ recovers \eqref{eq:maxsim}. $s_{\mathrm{tok}}$ is monotone in $r_i$. The argmax cluster is the same as in standard MaxSim, and Section~\ref{sec:loc} localizes with that cluster.

On the unit sphere in $d$ dimensions, a query token and a random centroid have an inner product of mean $0$ and variance $1/d$. The expected maximum over $K$ centroids is at most
\begin{equation}
\mathbb{E}\Big[\max_{1 \le j \le K} \langle \hat{q}_i, v_j \rangle\Big] \;\le\; \sqrt{\frac{2 \ln K}{d}} \;=\; \sqrt{\frac{2 \ln 128}{64}} \;\approx\; 0.389.
\label{eq:bound}
\end{equation}
An unrelated record therefore contributes about $0.389$ per query token. The floor is set to $\theta_{\mathrm{nf}}=0.38$. On ViDoRe V3 the measured mean of $r_i$ is $0.393$ on unrelated pages and $0.694$ on relevant pages, and $92\%$ of tokens on relevant pages exceed $0.38$. With $\beta_{\mathrm{nf}}=2.5$, a token at the relevant-page mean is scored about $3.8$ times the unrelated-page mean, while a match near the floor is almost unchanged. The noise floor is computed in S3 (Section~\ref{sec:retrieve}).

\section{Block-wise model merging}
\label{sec:merge}

\subsection{What is merged}

\begin{figure}[t]
\centering
\includegraphics[width=\linewidth]{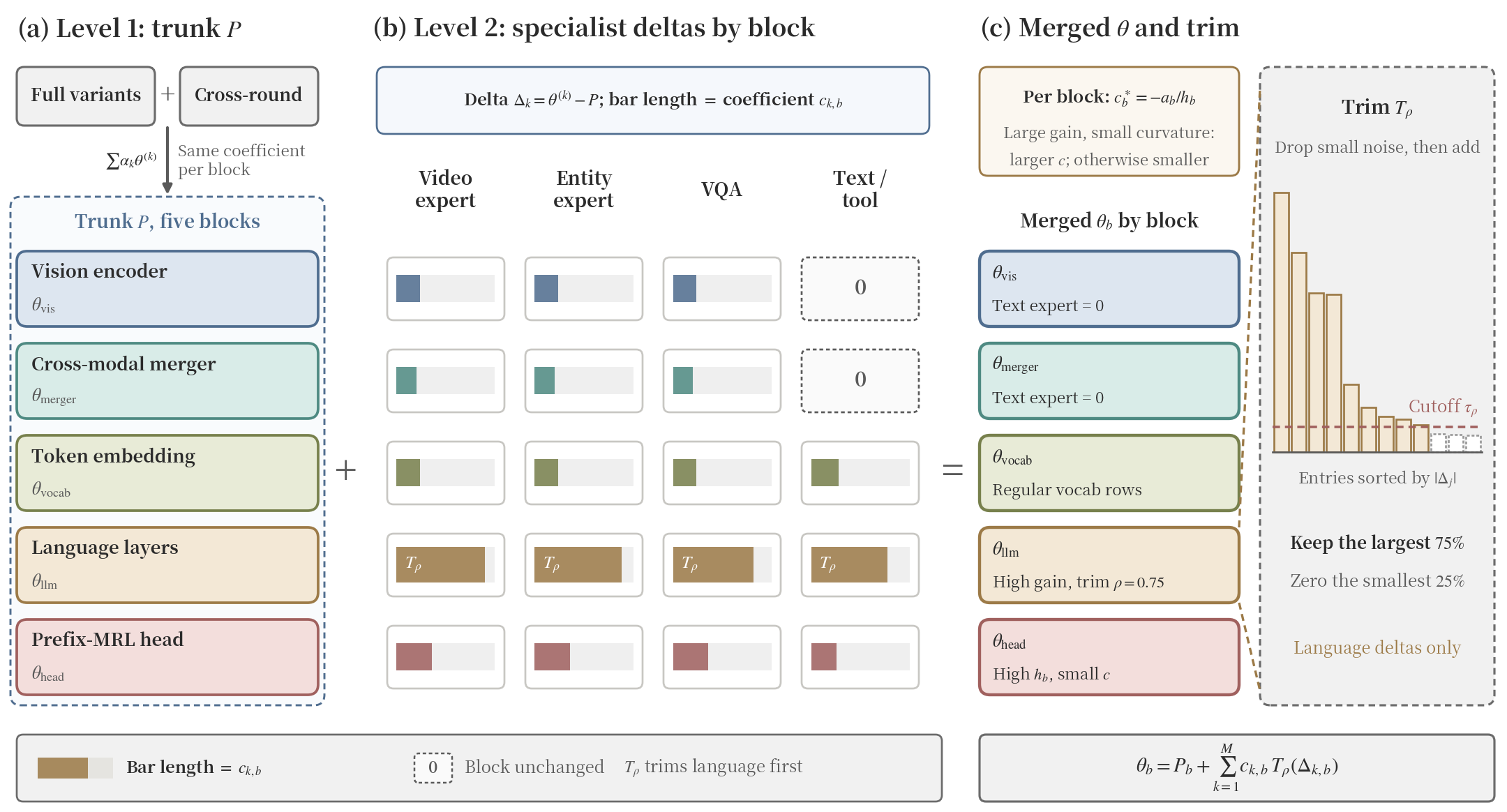}
\caption{Two-level block-wise merge. (a) Full-data variants from one start, and weights from neighboring rounds, are averaged with one coefficient per block into the trunk $P$. (b) Four specialists add $\Delta_k=\theta^{(k)}-P$, weighted per block; bar length is $c_{k,b}$. Text and tool specialists set the vision and merger coefficients to $0$, and language-layer deltas are trimmed by $\mathcal{T}_{\rho}$. (c) The per-block policy after the merge.}
\label{fig:merge}
\end{figure}

The weights are a merge of the models trained across stages. One level averages full-data runs that share a starting point, and averages neighboring stages. After stage 3, a second level trains specialists for video, fine-grained entities, visual question answering, and tool use, then adds their deltas back to the trunk (Figure~\ref{fig:merge}). Each ability is trained on its own data and added in parameter space \citep{ilharco2023taskarith}.

Parameters are split into five blocks $b \in \{\mathrm{vis},\mathrm{merger},\mathrm{vocab},\mathrm{llm},\mathrm{head}\}$: the vision encoder, the merger, the token embedding, the language layers, and the Prefix-MRL head. Let $P$ be the trunk and $\theta^{(k)}$ the $k$-th model. The delta is $\Delta_k = \theta^{(k)} - P$, and $\Delta_{k,b}$ is its restriction to block $b$. The merged block is
\begin{equation}
\theta_b = P_b + \sum_{k=1}^{M} c_{k,b}\, \mathcal{T}_{\rho}\!\left(\Delta_{k,b}\right),
\qquad c_{k,b} \in [0, 1].
\label{eq:merge}
\end{equation}
A uniform average is the special case in which every block shares one coefficient. A text or tool specialist takes coefficient $0$ on the vision encoder and the merger, and a positive coefficient on the language layers. In Figure~\ref{fig:merge} the length of a bar is that coefficient. The vision bar of a text or tool specialist is empty because the coefficient is zero.

$\mathcal{T}_{\rho}$ keeps the largest-magnitude fraction $\rho$ of each tensor and zeros the rest \citep{yadav2023ties}:
\begin{equation}
\mathcal{T}_{\rho}(\Delta)_{j} =
\begin{cases}
\Delta_{j}, & |\Delta_{j}| \ge \tau_{\rho}, \\
0, & \text{otherwise},
\end{cases}
\label{eq:trim}
\end{equation}
where $\tau_{\rho}$ is the $(1-\rho)$-quantile of $|\Delta|$. Most entries in a specialist delta are small and close to random. They add little to the target ability and interfere with other specialists. Language-layer deltas use $\rho=0.75$. Vision, merger, and head deltas use $\rho=1$.

\subsection{Per-block coefficients}

A second-order expansion writes the extra loss of one coefficient shared by every block. Consider a single model, with trimmed block delta $\delta_b$ and coefficient $c_b$. Let $g$ and $H$ be the gradient and Hessian of the loss at $P$, and ignore cross-block second derivatives. The loss after the merge is approximately
\begin{equation}
\mathcal{L}(\theta) \approx \mathcal{L}(P) + \sum_{b} \Big( a_{b}\, c_{b} + \frac{1}{2} h_{b}\, c_{b}^{2} \Big),
\qquad a_{b} = g_{b}^{\top}\delta_{b}, \quad h_{b} = \delta_{b}^{\top} H_{bb}\, \delta_{b}.
\label{eq:quad}
\end{equation}
$a_b<0$ means the delta reduces the loss, and $h_b>0$ is the curvature cost. Each block is a quadratic in $c_b$, with vertex
\begin{equation}
c_b^{\ast} = -\frac{a_b}{h_b},
\label{eq:vertex}
\end{equation}
and the coefficient used is that value clipped to $[0,1]$. If every block is forced to share a coefficient $c$, the best shared value is the curvature-weighted average of the per-block vertices, and the extra loss is
\begin{equation}
\Delta\mathcal{L} = \frac{1}{2} \sum_{b} h_{b} \big(\bar{c} - c_{b}^{\ast}\big)^{2} \;\ge\; 0.
\label{eq:extra}
\end{equation}
The gap grows as the per-block optima diverge. The projection head, which writes the stored vectors, has large curvature and a smaller optimal coefficient. The language layers carry most of the task ability and prefer a larger coefficient. A text or tool specialist has $a_b \ge 0$ on the vision encoder, so that coefficient is $0$. Trimming removes entries that barely change $a_b$ but inflate $h_b$ and the interference with other specialists. Measured effects of each merge step are in Table~\ref{tab:ablate}.

\section{ST-HAC}
\label{sec:sthac}

\subsection{Agglomerative clustering}

A high-resolution page or a $32$-frame clip emits hundreds to thousands of vectors, and they are not spread uniformly. The vectors that Section~1.3 counts as unselected are mostly collinear background and static frames \citep{qin2026compression}. The evidence that decides a retrieval, a single cell, a small logo, or a short action, is an isolated point far from the background. The compression merges collinear background and static frames into a few clusters, and leaves an isolated piece of evidence in a cluster of its own. Ward clustering merges the pair whose union increases the within-cluster sum of squares the least: nearby background tokens merge first, and an isolated point stays until the end. ST-HAC runs this clustering in a spatiotemporal feature space. Table~\ref{tab:compress} compares it with uniform pooling and random pruning.

\subsection{One procedure for every modality}

\begin{figure}[t]
\centering
\includegraphics[width=\linewidth]{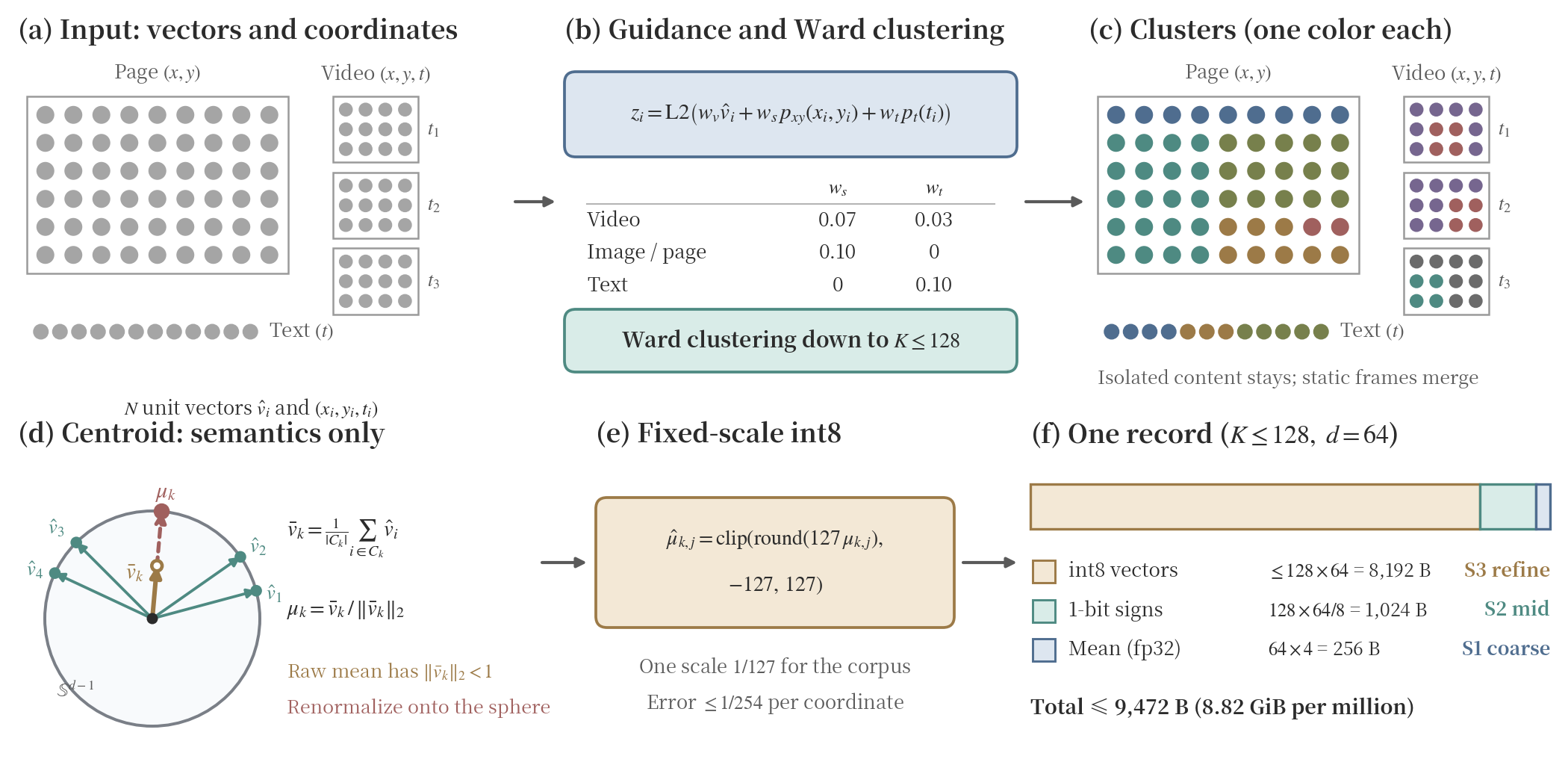}
\caption{ST-HAC offline compression. (a)(b) Ward clustering on semantic vectors plus the spatiotemporal code. (c) An isolated region stays its own cluster, and adjacent static frames merge. (d) A centroid is the renormalized mean of the semantic vectors only. (e) Fixed-scale int8. (f) The bytes in one record.}
\label{fig:sthac}
\end{figure}

After the $d=64$ projection and per-token $\ell_2$ normalization, any input is a sequence $\{\hat{v}_i\}_{i=1}^{N} \subset \mathbb{S}^{d-1}$. ST-HAC replaces that sequence by at most $K \le 128$ vectors of dimension $64$. Figure~\ref{fig:sthac} shows the steps. If clustering used semantics alone, similar digits in two columns of a report, or similar backgrounds far apart in a video, would fall into one cluster. Each token therefore receives normalized coordinates $(x_i,y_i,t_i)\in[0,1]^3$. For a video packed with stride $\Delta t_{\mathrm{patch}}=2$ into $F$ groups,
\begin{equation}
t_i = \frac{f_i}{\max(F-1,\, 1)}.
\label{eq:tcoord}
\end{equation}
For a text sequence, $t_i=(i-1)/\max(N-1,1)$. The clustering feature mixes the semantic vector with sinusoidal space and time codes and renormalizes:
\begin{equation}
z_i = \operatorname{L2}\!\Big(
w_v\,\hat{v}_i + w_s\,p_{xy}(x_i,y_i) + w_t\,p_t(t_i)
\Big),
\qquad w_v = 1-w_s-w_t.
\label{eq:guide}
\end{equation}
Video uses $(w_s,w_t)=(0.07,0.03)$. Images and pages use $(0.10, 0)$. Text uses $(0, 0.10)$. A missing axis has weight zero. The code and the output format are shared.

Long inputs are segmented first. Video is cut at $60$ seconds and text at $1{,}024$ tokens. Each segment is clustered, and the segment centroids are clustered again up to $K$ with a segment-time code of weight $w_t=0.03$. The Ward distance matrix then stays inside a segment. Ward clustering on $\{z_i\}$ stops at $K$ clusters. If $N\le K$, the full sequence is kept. Position weights are much smaller than the semantic weight, so the partition is still driven by semantics.

\subsection{A monotone time code}

The spatial code $p_{xy}$ is a standard multi-frequency sinusoid. The same frequencies are unsafe on time. A frequency above $\pi$ wraps, so two shots that are far apart can receive similar codes. ST-HAC draws $M=d/2$ frequencies evenly from $[\pi/4,\pi]$,
\begin{equation}
p_t(t) = \frac{1}{\sqrt{M}}
\big(\sin\omega_1 t,\ \cos\omega_1 t,\ \ldots,\ \sin\omega_M t,\ \cos\omega_M t\big)^{\!\top}.
\label{eq:pt}
\end{equation}
The leading factor is $1/\sqrt{M}$. Each sine--cosine pair then contributes $1/M$ to the squared norm, and $M$ pairs give $\|p_t(t)\|_2^{2}=1$. The inner product of two times separated by $\Delta t$ is
\begin{equation}
K_t(\Delta t) = \langle p_t(t),\, p_t(t+\Delta t)\rangle = \frac{1}{M}\sum_{m=1}^{M} \cos(\omega_m \Delta t).
\label{eq:kt}
\end{equation}
Every $\omega_m \Delta t$ for $\Delta t\in[0,1]$ lies in $[0,\pi]$, where cosine decreases, so $K_t$ decreases from $K_t(0)=1$. Adjacent static frames merge first, and distant shots stay apart. The time coordinate is normalized to $[0,1]$, independent of frame rate and duration.

\subsection{Semantic centroids}

The position features set the partition and then leave the index. Otherwise the inner product with a query would contain a content-independent bias. Given clusters $\mathcal{C}=\{C_1,\ldots,C_K\}$, the centroid uses only the original semantic vectors, $\bar{v}_k = \frac{1}{|C_k|}\sum_{i\in C_k}\hat{v}_i$, and is renormalized:
\begin{equation}
\mu_k = \frac{\bar{v}_k}{\|\bar{v}_k\|_{2}} \in \mathbb{S}^{d-1}.
\label{eq:centroid}
\end{equation}
Because every member is a unit vector,
\begin{equation}
\frac{1}{|C_k|}\sum_{i \in C_k} \|\hat{v}_i - \bar{v}_k\|_{2}^{2} = 1 - \|\bar{v}_k\|_{2}^{2}.
\label{eq:disp}
\end{equation}
A more dispersed cluster has a shorter mean. Scoring with $\bar{v}_k$ would penalize clusters that cover more content. After \eqref{eq:centroid} the inner product is $\cos\theta$, which matches the score used during training. Table~\ref{tab:compress} compares the two.

\subsection{Fixed-scale int8 and the bytes in one record}

Each coordinate of $\mu_k$ lies in $[-1,1]$, and the corpus shares the scale $1/127$. Indexing rounds at that scale, $\hat{\mu}_{k,j} = \operatorname{clip}(\operatorname{round}(127\,\mu_{k,j}), -127, 127)$.
The per-coordinate error is at most $1/254$. Scores accumulate in integer arithmetic, and a new record is appended directly. Table~\ref{tab:int8} compares the scales. At $K\le 128$ and $d=64$, the int8 multi-vectors take at most $8{,}192$ bytes, the fp32 mean takes $256$ bytes, and the optional $1$-bit signs take $1{,}024$ bytes. The two-level index is at most $8{,}448$ bytes ($7.87$ GiB per million records). With the sign bits, the three-stage index is at most $9{,}472$ bytes ($8.82$ GiB per million records).

\section{Three-stage retrieval}
\label{sec:retrieve}

\subsection{Three views of one record}

\begin{figure}[H]
\centering
\includegraphics[width=\linewidth]{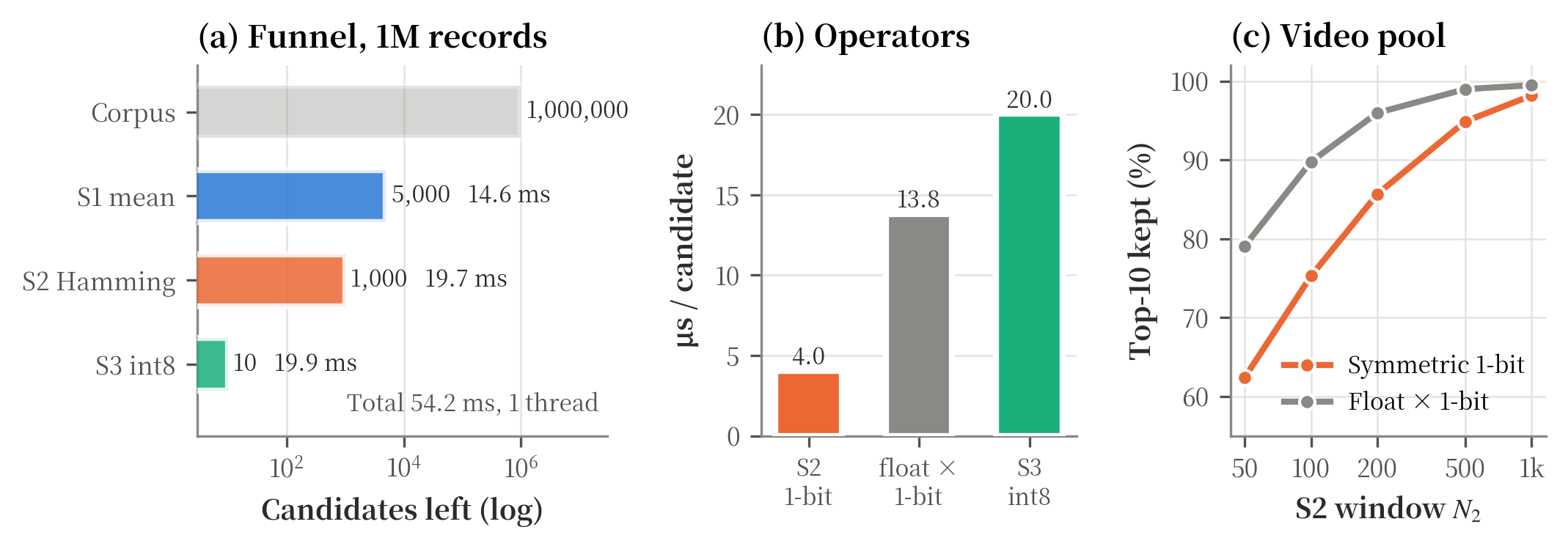}
\caption{Three-stage retrieval. (a) The candidate window and the time of each stage on one million records. (b) Time per candidate of the three operators, at the median query length and $K=128$. (c) The S2 window $N_2$ and top-10 retention on the mixed video pool.}
\label{fig:funnel}
\end{figure}

A two-stage pipeline often reads a full-precision side index after the coarse pass, and that side index is hundreds of gigabytes at million scale. OmniEVIE reads all three stages from the record in Section~\ref{sec:sthac} (Figure~\ref{fig:funnel}, Table~\ref{tab:stages3}).

\begin{table}[t]
\centering
\small
\setlength{\tabcolsep}{3pt}
\caption{Candidate representation, operator, and bytes of the three stages. $K'$ is the number of stored centroids, at most $128$.}
\label{tab:stages3}
\begin{tabular}{llllr}
\toprule
Stage & Representation & Operator & Window & Bytes \\
\midrule
S1 & $\mu_{\mathrm{avg}}=\operatorname{L2}(\frac{1}{K'}\sum_k \mu_k)$ & Exact inner product or ANN & $N_{\mathrm{db}}\to N_1$ & $256$ \\
S2 & $\operatorname{sign}(\mu_k)$ & Hamming MaxSim, XOR and popcount & $N_1\to N_2$ & $1{,}024$ \\
S3 & Fixed-scale int8 $\hat{\mu}_k$ & Int8 MaxSim, plus \eqref{eq:nf} & $N_2\to \mathrm{top}\text{-}k$ & $8{,}192$ \\
\bottomrule
\end{tabular}
\end{table}

The query is encoded once, and its mean and sign bits are read from that same output.

\subsection{Latency and window trade-offs}

With corpus size $N_{\mathrm{db}}$, $L_q$ query tokens, at most $K$ centroids, and dimension $d$, the retrieval time is
\begin{equation}
T_{\mathrm{total}} =
c_{\mathrm{s1}}\, N_{\mathrm{db}}\, d
+ c_{\mathrm{pop}}\, N_{1}\, L_{q}\, K
+ c_{\mathrm{int8}}\, N_{2}\, L_{q}\, K.
\label{eq:time}
\end{equation}
The first term is the S1 scan, the second is Hamming S2, and the third is int8 S3. At $d=64$ the sign pattern of one centroid is one $64$-bit word, scored by one XOR and one popcount, while the int8 dot product performs $64$ multiply-adds. At the median query length and $K=128$, int8 scoring costs about $20.0\,\mu\mathrm{s}$ per candidate and Hamming scoring about $4.0\,\mu\mathrm{s}$, so $r = c_{\mathrm{int8}}/c_{\mathrm{pop}} = 5$. Holding the sum of S2 and S3 fixed, dropping $\Delta N_2$ int8 candidates buys
\begin{equation}
\Delta N_1 = r\,\Delta N_2
\label{eq:trade}
\end{equation}
extra candidates in the S1 window. An S1 window that feeds S3 directly stays small, because the int8 score dominates the time. On a larger corpus that fixed window is a smaller fraction of the records, and the mean vector drops the true hit before S3 ever sees it. S2 cuts the int8 window to about a thousand candidates and, by \eqref{eq:trade}, buys about $r$ times as many candidates in S1 at a lower total time.

S2 stores signs on both sides, so one pair is one XOR and one popcount. A floating-point query against a $1$-bit document still costs on the order of an int8 dot product; that row is the last line of Table~\ref{tab:latency}. Symmetric $1$-bit is the middle stage, and symmetric int8 is the final stage.

\subsection{Order preservation of the sign stage}

S2 keeps the relevant candidates inside the top $N_2$. S3 produces the final order. Signs keep the orthant of each coordinate. For unit vectors $u,v$ and a random rotation $R$ \citep{charikar2002simhash},
\begin{equation}
\mathbb{E}\!\left[\frac{1}{d}\langle \operatorname{sign}(Ru),\, \operatorname{sign}(Rv)\rangle\right] = 1 - \frac{2\theta}{\pi},
\qquad \theta = \arccos\langle u, v\rangle.
\label{eq:hamming}
\end{equation}
Hamming similarity is monotone in cosine, so the middle stage keeps that order. OmniEVIE takes signs in the original coordinates. Retention of the full-scan top-$10$ is reported in Table~\ref{tab:latency}.

\subsection{PLAID and MUVERA}

PLAID clusters the corpus into a global codebook and stores residuals in an inverted index \citep{santhanam2022plaid}. MUVERA projects the multi-vectors of a record to one fixed-dimensional encoding and reuses a single-vector index \citep{dhulipala2024muvera}. Table~\ref{tab:engines} compares the three as systems that have to be updated. PLAID fits a corpus that can be reclustered. MUVERA fits a stack that already searches single vectors. OmniEVIE fits a corpus that keeps receiving records of different modalities.

\begin{table}[t]
\centering
\scriptsize
\setlength{\tabcolsep}{3pt}
\caption{Deployment properties of three multi-vector retrieval designs.}
\label{tab:engines}
\begin{tabular}{p{0.18\linewidth}p{0.26\linewidth}p{0.24\linewidth}p{0.26\linewidth}}
\toprule
 & PLAID & MUVERA & OmniEVIE \\
\midrule
Index & Global codebook and residual postings & One high-dimensional FDE vector & Mean, $1$-bit signs, and int8 centroids on the record \\
Updates & Codebook is rebuilt when the corpus shifts & Fixed random hyperplanes & Append an independently encoded record \\
Filters & Scalar filters inside the postings & Filters of a single-vector engine & Restrict the S1 candidate set \\
Local hit & Centroid map, residuals must be decoded & Dropped by the projection & Kept, from the S3 argmax \\
Engine & Inverted index and residual decode & High-dimensional MIPS & Inner product, popcount, int8 dot product \\
\bottomrule
\end{tabular}
\end{table}

\subsection{Localizing the hit}
\label{sec:loc}

Each centroid corresponds to a set of patches or a time interval. Bounding boxes and time spans are metadata of the source document, and the S3 argmax names the hit.

For a page or an image, the metadata of $\mu_k$ stores the minimum rectangle of its member patches and the page index. For a high-weight query token, $\hat{k}=\arg\max_k \langle \hat{q}_i, \hat{\mu}_k\rangle$ highlights that rectangle. The argmax is an intermediate of MaxSim. Localization keeps a cluster whose rectangle covers at most $5\%$ of the page. A larger rectangle is whitespace and background spread over the page. For a long video, clips are encoded in fixed windows with the same record format, and each centroid stores $[t_{\min}, t_{\max}]$. The windows of one video are placed on a single timeline by video identifier, and the span is the heat of \eqref{eq:heat}.

The interval hit by one action word covers part of the segment, and the whole segment comes from placing every query token on one time axis. On the QVHighlights validation set, $60.7\%$ of the relevant segments cross a window boundary. Windows are encoded and clustered separately, and a centroid stays inside its window, so each token takes its best centroid in every window and evidence above the noise floor enters the heat on both sides of the boundary. Let $r_i^{(j)}$ be the maximum similarity of token $i$ inside window $j$ and $\hat{k}_i^{(j)}$ the corresponding centroid. The heat of group $g$ is
\begin{equation}
h(g) = \sum_{j=1}^{N_{\mathrm{win}}} \;\sum_{i:\, g \in [t_{\min}(\hat{k}_i^{(j)}),\, t_{\max}(\hat{k}_i^{(j)})]} w_i \operatorname{ReLU}\!\big(r_i^{(j)} - \theta_{\mathrm{nf}}\big).
\label{eq:heat}
\end{equation}
Each token adds the part of \eqref{eq:nf} above the noise floor, multiplied by the gate weight of \eqref{eq:gate}, onto the temporal groups of its centroid, and a larger weight contributes more. The group with the highest heat is the located point, and the connected segment where the heat is at least $80\%$ of the peak is the returned span. The ranking score still takes the global maximum of \eqref{eq:nf}. Figure~\ref{fig:loc} gives the examples, the measurements, and the context passed to the downstream generator.

\section{Experiments}

\subsection{Protocol}

Unless a row says otherwise, every OmniEVIE number uses $K\le 128$, $d=64$, and the fixed int8 scale. Task definitions follow MMEB \citep{jiang2024vlm2vec,meng2025vlm2vec,huang2026mmebv3}; Appendix~\ref{app:suites} lists the suites. MMEB-V2 numbers for WeMM-Embedding and Qwen3-VL-Embedding are the official leaderboard results. The MMEB-V3 text and agent numbers for Qwen3-VL-Embedding are taken from \citet{zhou2026wemm}. In Table~\ref{tab:vidore} the Qwen3-VL-Embedding-8B average is the published ViDoRe V3 number \citep{loison2026vidore}; the per-domain numbers are our rerun of that setup, and the rerun average is $58.37$. WeMM-Embedding-9B on MMEB-V3 is the leaderboard value $59.33$.

\subsection{MMEB-V2 and MMEB-V3}

Table~\ref{tab:mmebv3} breaks out the text and agent subgroups of MMEB-V3 (190 tasks). Text subgroups are reasoning retrieval (T-RR), instruction following (T-IF), long context (T-LC), multi-condition (T-MC), and general retrieval (T-GR). Agent subgroups are Tool, GUI, Memory, and MCMR. Table~\ref{tab:mmebv2} reports MMEB-V2 (78 tasks) as four families: all, image, video, and visual documents. The full-sequence rows are the blue line in Figure~\ref{fig:overview}.

\begin{table}[t]
\centering
\scriptsize
\setlength{\tabcolsep}{2.2pt}
\caption{MMEB-V3 over 190 tasks. All is the unweighted mean in Appendix~\ref{app:suites}. Text reports nDCG@5; other groups report Hit@1. The 8-frame row changes only the video frame cap. Bold marks the best value in the column.}
\label{tab:mmebv3}
\resizebox{\linewidth}{!}{%
\begin{tabular}{lrrrrrrrrrrrr}
\toprule
Model & All & Text & T-RR & T-IF & T-LC & T-MC & T-GR & Agent & Tool & GUI & Mem. & MCMR \\
\midrule
Tasks & 190 & 53 & 20 & 9 & 6 & 5 & 13 & 47 & 35 & 8 & 4 & 1 \\
Qwen3-VL-Emb-2B & 50.9 & 39.2 & 16.6 & 40.6 & 53.1 & 61.2 & 58.0 & 39.3 & 42.6 & 30.4 & 28.4 & 42.0 \\
Qwen3-VL-Emb-8B & 53.5 & 42.5 & 18.2 & 44.8 & 58.0 & 61.2 & 61.2 & 38.4 & 41.3 & 33.5 & 22.8 & 38.0 \\
WeMM-Emb-2B & 55.77 & 45.34 & 26.29 & 49.30 & 61.90 & 59.72 & 58.74 & 44.05 & 46.23 & 38.87 & 35.38 & 42.51 \\
WeMM-Emb-4B & 57.97 & 47.88 & 30.01 & 52.02 & 62.43 & 60.14 & 61.09 & 48.06 & 50.58 & 41.44 & \textbf{39.19} & 41.91 \\
WeMM-Emb-9B & 59.33 & 48.76 & \textbf{31.82} & 52.78 & 64.48 & 58.28 & 61.12 & 50.10 & 52.95 & \textbf{43.33} & 38.65 & 49.34 \\
OmniEVIE-4B & 60.06 & 51.77 & 26.36 & 69.65 & 70.71 & 67.75 & 63.58 & 54.03 & \textbf{58.84} & 41.34 & 37.35 & 47.39 \\
OmniEVIE-4B, 8 frames & 59.76 & 51.77 & 26.36 & 69.65 & 70.71 & 67.75 & 63.58 & 54.03 & 58.84 & 41.34 & 37.35 & 47.39 \\
OmniEVIE-4B, full sequence $d=2048$ & \textbf{60.68} & \textbf{52.97} & 27.46 & \textbf{70.81} & \textbf{73.04} & \textbf{68.55} & \textbf{64.60} & \textbf{54.14} & 58.81 & 42.00 & 37.60 & \textbf{49.74} \\
\bottomrule
\end{tabular}}
\end{table}

\begin{table}[t]
\centering
\small
\setlength{\tabcolsep}{6pt}
\caption{MMEB-V2 over 78 tasks. All is the unweighted mean. Image and video report Hit@1; visual documents report nDCG@5. WeMM-Emb-4B uses the same $4.5$B backbone as OmniEVIE. Bold marks the best value in the column.}
\label{tab:mmebv2}
\begin{tabular}{lrrrr}
\toprule
Model & All & Image & Video & VisDoc \\
\midrule
Tasks & 78 & 36 & 18 & 24 \\
Qwen3-VL-Emb-2B & 73.25 & 74.96 & 61.87 & 79.22 \\
Qwen3-VL-Emb-8B & 77.82 & 80.12 & 67.15 & 82.36 \\
WeMM-Emb-2B & 77.94 & 79.63 & 70.82 & 80.75 \\
WeMM-Emb-4B & 79.19 & 80.84 & 72.11 & 82.04 \\
WeMM-Emb-9B & \textbf{80.58} & \textbf{81.87} & \textbf{74.31} & 83.34 \\
OmniEVIE-4B & 77.96 & 78.09 & 69.62 & 84.02 \\
OmniEVIE-4B, 8 frames & 77.24 & 78.09 & 66.49 & 84.02 \\
OmniEVIE-4B, full sequence $d=2048$ & 78.56 & 78.70 & 70.08 & \textbf{84.70} \\
\bottomrule
\end{tabular}
\end{table}

OmniEVIE reaches $60.06$ on MMEB-V3. With video capped at the official $8$ frames the same weights score $59.76$, above the WeMM-Embedding-9B value of $59.33$. MMEB-V2 is $77.96$, which is $2.62$ below WeMM-Embedding-9B overall, $3.78$ on images, and $4.69$ on video. Those image and video gaps are in fine-grained classification and video retrieval. The MMEB-V3 gain is on long text: text overall is $51.77$, with instruction following at $69.65$, long context at $70.71$, and multi-condition retrieval at $67.75$. Agent overall is $54.03$, and tool retrieval is $58.84$. Late interaction matches each query condition to a different span of that text. The released index scores $84.02$ on visual documents. The highest entry in that column of Table~\ref{tab:mmebv2} is the full sequence at $d=2048$, $84.70$.

\subsection{Visual documents, fine-grained images, and video}

Table~\ref{tab:vidore} and Table~\ref{tab:fine} report visual documents, fine-grained images, and video.

\begin{table}[t]
\centering
\scriptsize
\setlength{\tabcolsep}{2.4pt}
\caption{ViDoRe V3 graded nDCG@10, macro-averaged over eight domains and six languages. Storage is GiB per million pages for the compressed rows. Bold marks the best value in the column.}
\label{tab:vidore}
\resizebox{\linewidth}{!}{%
\begin{tabular}{llrrrrrrrrr}
\toprule
Model & Index & Avg & Comp. & Energy & Fin-EN & Fin-FR & HR & Ind. & Med. & Phys. \\
\midrule
Nemotron-ColEmbed-V2-8B & Full multi-vector & 63.76 & 79.35 & 70.05 & 67.72 & 51.50 & \textbf{66.92} & 56.81 & 67.39 & 50.31 \\
Qwen3-VL-Emb-8B & Single vector & 59.00 & 75.86 & 64.47 & 60.88 & 44.70 & 58.88 & 47.78 & 66.05 & 48.31 \\
OmniEVIE-4B, $d=2048$ & Prefix-MRL, full sequence & \textbf{64.34} & \textbf{79.89} & \textbf{71.25} & \textbf{67.76} & \textbf{53.45} & 66.33 & \textbf{57.03} & \textbf{68.11} & \textbf{50.87} \\
OmniEVIE-4B, $K128\times d64$ & $7.87$ GiB/M & 62.96 & 78.73 & 69.94 & 66.01 & 52.52 & 64.68 & 55.08 & 66.47 & 50.28 \\
OmniEVIE-4B, $K32\times d64$ & $2.15$ GiB/M & 61.16 & 78.13 & 68.61 & 64.52 & 49.43 & 60.90 & 51.73 & 65.73 & 50.22 \\
\bottomrule
\end{tabular}}
\end{table}

\begin{table}[t]
\centering
\scriptsize
\setlength{\tabcolsep}{2.5pt}
\caption{Fine-grained image Top-1, video nDCG@10, and forward-versus-reversed accuracy. The four-set mean is MSR-VTT, MSVD, DiDeMo, and VATEX. Bold marks the best value in the column.}
\label{tab:fine}
\resizebox{\linewidth}{!}{%
\begin{tabular}{lrrrrrrrrrr}
\toprule
Model & FORB & S2S & MFashion & MSR & MSVD & DiDeMo & VATEX & Vid-4 & Order \\
\midrule
Qwen3-VL-Emb-8B & \textbf{92.80} & 62.90 & 65.10 & 71.95 & 77.84 & 74.54 & \textbf{76.60} & 75.23 & 51.63 \\
Frame-independent & -- & -- & -- & 63.76 & 73.11 & 78.30 & 69.37 & 71.13 & 51.17 \\
OmniEVIE-4B & 89.94 & \textbf{72.16} & \textbf{75.98} & \textbf{72.38} & \textbf{78.89} & \textbf{87.54} & 76.14 & \textbf{78.74} & \textbf{75.63} \\
\bottomrule
\end{tabular}}
\end{table}

In Table~\ref{tab:vidore}, the $K128\times d64$ int8 index scores $62.96$ on ViDoRe V3, against $64.34$ for the uncompressed $d=2048$ sequence. The compressed index is $7.87$ GiB per million pages, about $1/486$ of the fp16 full sequence, and the loss is $1.38$ points. Reducing $K$ further to $32$, or $2.15$ GiB per million pages, leaves $61.16$. Computing and energy are the highest domains ($78.73$ and $69.94$); French finance and physics are the lowest ($52.52$ and $50.28$).

In Table~\ref{tab:fine}, Top-1 on FORB, Street2Shop, and MovingFashion is $89.94$, $72.16$, and $75.98$. Street2Shop and MovingFashion pair product photos taken in different scenes, so the match is a local part of the product across different scenes. OmniEVIE averages $78.74$ across the four video datasets. Replacing the native video tokens by an independent encoding of each frame drops that average to $71.13$ and drops forward-versus-reversed accuracy from $75.63\%$ to $51.17\%$. The forward and reversed clips then produce the same set of vectors, and the accuracy falls to chance.

\subsection{Ablations}

In the M rows of Table~\ref{tab:ablate}, each cell is the score before the merge and the score after it. The video number in T1 uses $8$ frames.

\begin{table}[t]
\centering
\scriptsize
\setlength{\tabcolsep}{3pt}
\caption{Stage, query-side, and merge ablations on MMEB-V3. M cells are written as before $\to$ after. T1 video uses 8 frames.}
\label{tab:ablate}
\begin{tabular}{llrrrrrr}
\toprule
ID & Variant & V3 & Image & Video & VisDoc & Text & Agent \\
\midrule
T1 & End of stage 2 & 44.38 & 39.33 & 45.17 & 83.59 & 45.74 & 36.73 \\
T2 & End of stage 3 & 58.14 & 75.67 & 67.25 & 82.71 & 50.77 & 50.85 \\
T3 & Final model & 60.06 & 78.09 & 69.62 & 84.02 & 51.77 & 54.03 \\
Q1 & Gate only & 59.85 & 78.02 & 69.65 & 83.87 & 51.71 & 53.35 \\
Q2 & Uniform MaxSim & 59.71 & 78.04 & 69.37 & 83.80 & 51.64 & 52.99 \\
M1 & One variant $\to$ linear average & 58.37$\to$58.80 & 76.34$\to$76.97 & 67.90$\to$67.81 & 82.93$\to$83.25 & 50.96$\to$51.27 & 50.65$\to$51.41 \\
M2 & One round $\to$ cross-round & 59.16$\to$59.29 & 77.15$\to$77.59 & 68.74$\to$68.76 & 83.69$\to$83.83 & 51.24$\to$51.30 & 52.25$\to$52.31 \\
M3 & Trunk $\to$ four specialists & 59.44$\to$59.67 & 77.84$\to$78.11 & 68.95$\to$69.31 & 83.98$\to$83.66 & 51.48$\to$51.67 & 52.34$\to$52.83 \\
M4 & Shared coefficient $\to$ per block & 59.60$\to$59.66 & 77.95$\to$77.98 & 69.24$\to$69.36 & 83.84$\to$83.62 & 51.56$\to$51.65 & 52.72$\to$52.94 \\
\bottomrule
\end{tabular}
\end{table}

Stage 3 is where image, text, and agent move. Adding classification, question answering, and tools takes image from $39.33$ to $75.67$, text from $45.74$ to $50.77$, and agent from $36.73$ to $50.85$, while visual documents stay near $83$. Against uniform MaxSim, the gate adds $0.36$ on agent and $0.28$ on video; the noise floor then adds $0.68$ on agent and $0.15$ on visual documents. Tool and memory candidates are long documents, and a few query tokens decide the score; amplifying the excess above the floor gives those tokens a larger share. Each of the four merge steps operates by parameter addition. Averaging full-data runs from one start adds $0.76$ on agent and $0.63$ on image. Merging across rounds adds $0.44$ on image. The four specialists add $0.49$ on agent, $0.36$ on video, and $0.27$ on image, and visual documents change by $-0.32$. A per-block coefficient, compared with one shared coefficient, adds $0.22$ on agent and $0.12$ on video, as \eqref{eq:extra} predicts, and visual documents change by $-0.22$.

\subsection{Compression methods at the same budget}

\begin{table}[t]
\centering
\small
\setlength{\tabcolsep}{4pt}
\caption{Compression at $K=128$, $d=64$, scored by ViDoRe V3 nDCG@10. Indexing time is single-thread milliseconds per page.}
\label{tab:compress}
\begin{tabular}{lcrr}
\toprule
Method & Training & nDCG@10 & ms / rec. \\
\midrule
Full sequence & -- & 63.09 & -- \\
Uniform grid pool & At indexing & 56.50 & 2.3 \\
Keep $K$ random tokens & At indexing & 60.41 & 0.1 \\
ST-HAC & With training & 62.96 & 20.0 \\
ST-HAC, $w_s=0$ & With training & 62.90 & 20.1 \\
ST-HAC, raw mean & With training & 63.00 & 20.0 \\
\bottomrule
\end{tabular}
\end{table}

At the same budget of $128$ int8 vectors of dimension $64$, ST-HAC scores $62.96$, the full sequence scores $63.09$, uniform pooling scores $56.50$, and random token retention scores $60.41$. Two ablations stay within $0.06$ of ST-HAC: $w_s=0$ drops the spatial term from the clustering feature and groups by semantics and time alone ($62.90$), and the raw-mean variant skips the centroid renormalization of \eqref{eq:centroid} and keeps the raw norm of the cluster mean ($63.00$). Building the index takes about $20.0$ single-thread milliseconds per page. Records are split across processes: one hundred million pages take about $2\times10^{6}$ single-thread seconds, or about $9$ hours over $64$ processes, and a new record is appended directly.

\subsection{Business scenarios}

Beyond the public benchmarks, we evaluate on real business scenarios (finance, industry, museums, landmarks, and automotive) covering text-to-image, image-to-image, and combined image-text retrieval; the scenario names are generalized and do not map to specific data sources. Table~\ref{tab:biz} reports Top-1 against two strong business baselines, Qwen3-VL-Embedding-8B and WeMM-Embedding-9B.

\begin{table}[t]
\centering
\small
\setlength{\tabcolsep}{5pt}
\caption{Business-scenario retrieval, Top-1, averaged within each scenario. Scenario names are generalized. Bold marks the best value in the row.}
\label{tab:biz}
\begin{tabular}{lrrr}
\toprule
Scenario & Qwen3-VL-Emb-8B & WeMM-Emb-9B & OmniEVIE-4B \\
\midrule
Text to image (mean) & 54.6 & 61.2 & \textbf{64.2} \\
Image to image (mean) & 93.4 & 94.1 & \textbf{96.5} \\
Image-text (mean) & 81.2 & 84.1 & \textbf{90.4} \\
Finance, text to image & \textbf{76.3} & 66.4 & \textbf{76.3} \\
Industry, text to image & 61.8 & 70.9 & \textbf{81.8} \\
Museums, text to image & 53.2 & 67.5 & \textbf{79.2} \\
Landmarks, image to image & 78.6 & 81.4 & \textbf{92.9} \\
Automotive, image-text & 86.4 & 90.9 & \textbf{100.0} \\
\bottomrule
\end{tabular}
\end{table}

On the category means OmniEVIE is first on all three: text-to-image $64.2$ against $61.2$, image-to-image $96.5$ against $94.1$, and image-text $90.4$ against $84.1$. Among representative scenarios, finance, industry, and museums in text-to-image reach $76.3$, $81.8$, and $79.2$, which are $9.9$, $10.9$, and $11.7$ points above the $9$B baseline; landmarks in image-to-image reach $92.9$, a gain of $11.5$; and automotive in image-text reaches $100.0$, a gain of $9.1$. The gains are largest where matching needs layout detail or a local product feature, consistent with the public-set results above.

\section{Analysis}

\subsection{Two degrees of freedom in the byte budget}

At a fixed number of bytes the index can keep more vectors $K$ or a higher dimension $d$. Table~\ref{tab:budget} separates the two choices. The first rows are the full Prefix-MRL sequence at each width. The last comparison holds the multi-vector payload at $4{,}096$ bytes and asks whether $K=64, d=64$ beats $K=32, d=128$. Section~1 says the count decides how many local pieces can be matched at once, while the dimension decides how well one vector separates lookalikes. When the evidence is scattered, the count should matter more.

\begin{table}[t]
\centering
\small
\setlength{\tabcolsep}{4pt}
\caption{Prefix-MRL width and ST-HAC budget on ViDoRe V3. Bytes include the mean vector where ST-HAC stores one. Storage is per million records.}
\label{tab:budget}
\begin{tabular}{lrrrr}
\toprule
Setting & $(K, d)$ & Bytes / rec. & GiB / M & ViDoRe V3 \\
\midrule
Prefix-MRL, full & seq.\ $\times 2048$ & $4{,}106{,}184$ & $3{,}824$ & 64.34 \\
Prefix-MRL, full & seq.\ $\times 512$ & $1{,}026{,}546$ & $956$ & 64.12 \\
Prefix-MRL, full & seq.\ $\times 128$ & $256{,}636$ & $239$ & 63.55 \\
Prefix-MRL, full & seq.\ $\times 64$ & $128{,}318$ & $120$ & 63.09 \\
ST-HAC & $\le 128 \times 64$ & $8{,}448$ & $7.87$ & 62.96 \\
ST-HAC & $\le 64 \times 64$ & $4{,}352$ & $4.05$ & 62.47 \\
ST-HAC & $\le 32 \times 128$ & $4{,}608$ & $4.29$ & 61.82 \\
ST-HAC & $\le 32 \times 64$ & $2{,}304$ & $2.15$ & 61.16 \\
\bottomrule
\end{tabular}
\end{table}

\begin{figure}[t]
\centering
\includegraphics[width=0.78\linewidth]{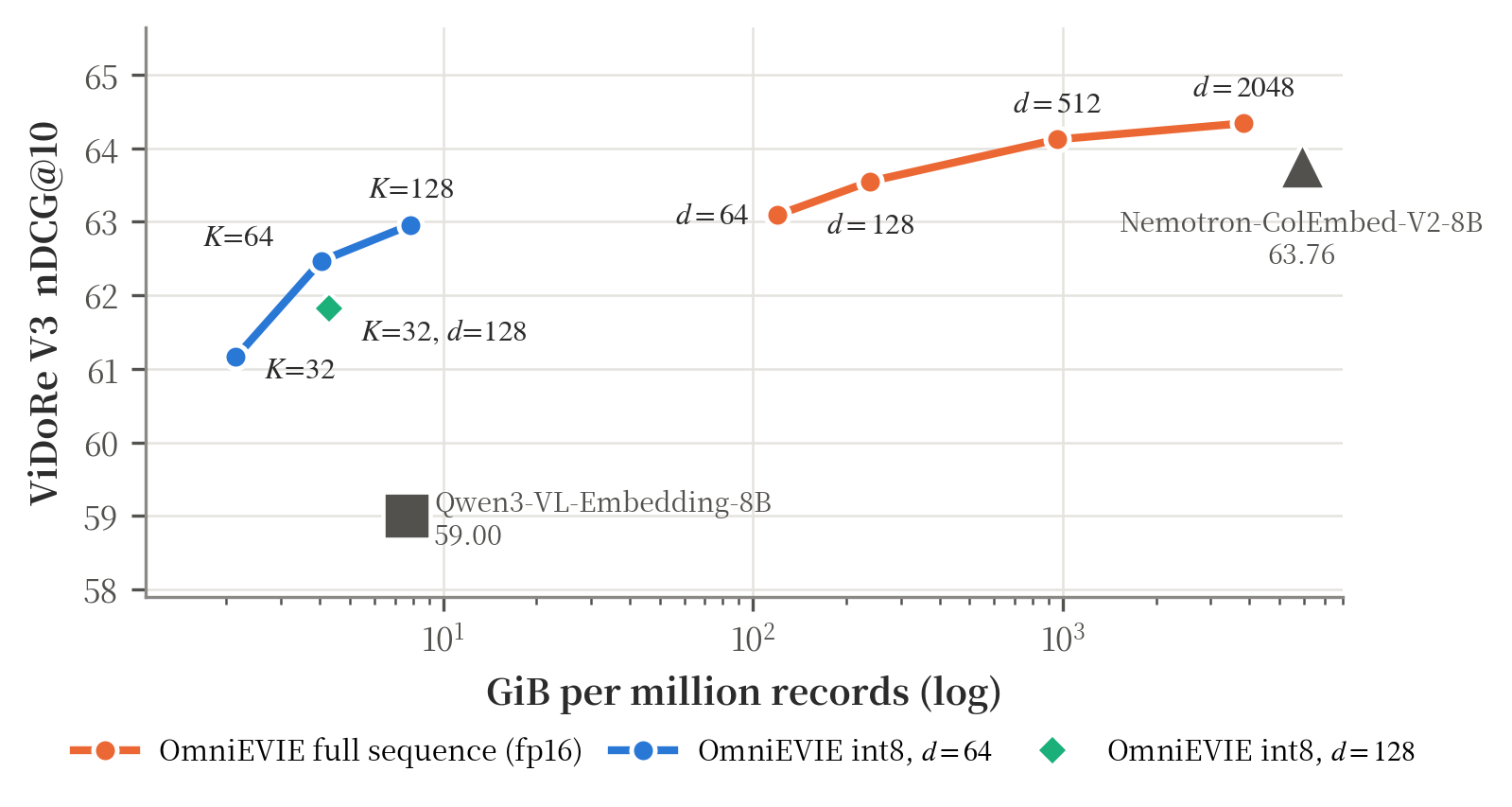}
\caption{Storage against ViDoRe V3 accuracy for the Prefix-MRL width and the ST-HAC budget, per million records on a log axis. The gray square and triangle are the single-vector baseline Qwen3-VL-Embedding-8B and the multi-vector baseline Nemotron-ColEmbed-V2-8B, placed at their own index storage (the latter at $4096$ dimensions and about $773$ tokens per page, about $5{,}898$ GiB per million in fp16); marker size reflects the roughly $8$B model scale, larger than the $4.5$B of OmniEVIE.}
\label{fig:budget}
\end{figure}

As shown in Figure~\ref{fig:budget} and Table~\ref{tab:budget}, cutting the full sequence from dimension $2048$ to $512$, $128$, and $64$ yields $64.34$, $64.12$, $63.55$, and $63.09$. A $32\times$ cut in storage costs $1.25$ points, so most of the full-sequence score survives a much narrower vector. At a payload of $4{,}096$ bytes, $K=64, d=64$ scores $62.47$ and $K=32, d=128$ scores $61.82$. Spending the bytes on more vectors beats spending them on a higher dimension.

On MMEB the same comparison uses the full $d=2048$ sequence, the blue line in Figure~\ref{fig:overview}, as the 190-task mean in Appendix~\ref{app:suites} (Table~\ref{tab:ceiling}). The released index scores $60.06$ on V3, $0.62$ below the ceiling of $60.68$, and $77.96$ on V2, $0.60$ below $78.56$. A full sequence cut to $d=64$ scores $59.02$ on V3, below the released index.

\begin{table}[t]
\centering
\small
\setlength{\tabcolsep}{4pt}
\caption{Full-sequence ceiling and the released index on MMEB. V3 and V2 use the means in Appendix~\ref{app:suites}.}
\label{tab:ceiling}
\begin{tabular}{lrrrrrrr}
\toprule
Representation & V3 & V2 & Image & Video & VisDoc & Text & Agent \\
\midrule
Full sequence, $d=2048$ & \textbf{60.68} & \textbf{78.56} & \textbf{78.70} & \textbf{70.08} & \textbf{84.70} & \textbf{52.97} & \textbf{54.14} \\
Full sequence, $d=64$ & 59.02 & 77.89 & 78.06 & 69.52 & 83.90 & 52.50 & 49.17 \\
$K128 \times d64$ int8 & 59.71 & 77.81 & 78.04 & 69.37 & 83.80 & 51.64 & 52.99 \\
Released index & 60.06 & 77.96 & 78.09 & 69.62 & 84.02 & 51.77 & 54.03 \\
\bottomrule
\end{tabular}
\end{table}

\subsection{Int8 scales}

Table~\ref{tab:int8} compares the fixed scale of Section~\ref{sec:sthac} on ViDoRe V3.

\begin{table}[t]
\centering
\small
\setlength{\tabcolsep}{4pt}
\caption{Int8 quantization on ViDoRe V3, graded nDCG@10.}
\label{tab:int8}
\begin{tabular}{lcccr}
\toprule
Scheme & Extra / vec. & Compare & Append & nDCG@10 \\
\midrule
fp16 centroids & $+64$ B & Float & Direct & 62.99 \\
int8 + fp16 scale & $+2$ B & Float & Direct & 62.96 \\
int8, scale dropped & $0$ & Integer & Direct & 57.82 \\
Corpus max$/127$ & $0$ & Integer & Rescan & 62.97 \\
Fixed scale $1/127$ & $0$ & Integer & Direct & 62.96 \\
\bottomrule
\end{tabular}
\end{table}

The fixed scale scores $62.96$, within $0.03$ of fp16 centroids ($62.99$), a per-vector fp16 scale ($62.96$), and a corpus-wide scale ($62.97$). Dropping the scale falls to $57.82$.

\subsection{Latency}

Table~\ref{tab:latency} reports single-thread CPU retrieval time at three corpus sizes. The operating point is $N_1=5{,}000$ and $N_2=1{,}000$. Deltas and top-10 retention are measured against full int8 MaxSim on a ViDoRe V3 pool of $19{,}252$ pages and a video pool of $13{,}932$ clips. Table~\ref{tab:qlen} varies the query length on the one-million corpus.

\begin{table}[t]
\centering
\small
\setlength{\tabcolsep}{3pt}
\caption{Single-thread retrieval latency in milliseconds. Cells are totals. $\Delta$ is the nDCG@10 change on the document pool and the video pool.}
\label{tab:latency}
\begin{tabular}{lrrrrrr}
\toprule
Pipeline & $N_1$ & $N_2$ & 100k & 1M & 5M & $\Delta$ doc / vid \\
\midrule
Full int8 MaxSim & all & -- & $1{,}657$ & $12{,}781$ & $63{,}892$ & $0$ / $0$ \\
Single vector $d=2048$ & all & -- & 34.6 & 331.0 & $1{,}666$ & -- \\
Single vector $d=4096$ & all & -- & 60.9 & 604.0 & $3{,}039$ & -- \\
S1 + S3 & $2{,}000$ & -- & 41.8 & 55.2 & 128.5 & $-0.60$ / $-0.66$ \\
S1 + S3 & $5{,}000$ & -- & 103.9 & 115.1 & 186.5 & $-0.13$ / $-0.10$ \\
Three-stage & $5{,}000$ & $200$ & 24.6 & 38.4 & 110.2 & $-0.20$ / $-3.09$ \\
Three-stage & $5{,}000$ & $500$ & 30.6 & 44.2 & 116.6 & $-0.13$ / $-0.84$ \\
Three-stage & $5{,}000$ & $1{,}000$ & 41.9 & 54.2 & 120.2 & $-0.13$ / $-0.26$ \\
Three-stage & $10{,}000$ & $1{,}000$ & 62.3 & 75.7 & 127.7 & $0.00$ / $-0.35$ \\
Float $\times$ $1$-bit & $5{,}000$ & $100$ & 72.7 & 85.8 & 156.7 & $-0.21$ / $-1.05$ \\
\bottomrule
\end{tabular}
\end{table}

Top-10 retention for the operating point ($N_1=5{,}000$, $N_2=1{,}000$) is $99.5\%$ on documents and $98.2\%$ on video. The same windows at $N_2=200$ retain $98.5\%$ and $85.7\%$. At $N_2=500$ they retain $99.4\%$ and $94.9\%$. At $N_1=10{,}000$ they retain $99.9\%$ and $97.7\%$. The two-stage rows at $N_1=2{,}000$ and $N_1=5{,}000$ retain $97.8\%/97.8\%$ and $99.5\%/99.6\%$. The asymmetric float$\times$$1$-bit row retains $98.6\%$ and $89.8\%$.

\begin{table}[t]
\centering
\small
\setlength{\tabcolsep}{4pt}
\caption{Latency in milliseconds by query length on $1$ million records, at $N_1=5{,}000$ and $N_2=1{,}000$. Speedup is against full int8 MaxSim.}
\label{tab:qlen}
\begin{tabular}{lrrrrr}
\toprule
Query & $L_q$ & S1 & S2 & S3 & Total \\
\midrule
Short text & $20$--$40$ & 14.8 & 14.6 & 17.3 & 46.7 ($274\times$) \\
Image or image-text & $90$--$270$ & 14.8 & 91.6 & 49.8 & 156.2 ($82\times$) \\
Long context or a full page & $150$--$500+$ & 14.8 & 321.8 & 103.3 & 439.9 ($29\times$) \\
\bottomrule
\end{tabular}
\end{table}

On one million records the operating point takes $54.2$ ms, of which S1, S2, and S3 take $14.6$, $19.7$, and $19.9$ ms. Full int8 MaxSim takes $12{,}781$ ms, so the three-stage path is $236\times$ faster, and nDCG@10 changes by $-0.13$ on documents and $-0.26$ on video. The same split is $1.5$ / $19.6$ / $20.8$ ms at $100$ thousand records. At five million records S1 grows to $80.5$ ms, because it scans the whole corpus, and the total is $120.2$ ms.

S1 is a $64$-dimensional scan, so the same comparison also favors OmniEVIE over an exact high-dimensional single vector once the corpus is large. At $100$ thousand records, $d=2048$ and $d=4096$ take $34.6$ ms and $60.9$ ms, close to the three-stage total. At one million records they take $331.0$ ms and $604.0$ ms, against $54.2$ ms ($6.1\times$ and $11.1\times$). At five million records they take $1{,}666$ ms and $3{,}039$ ms, against $120.2$ ms ($13.9\times$ and $25.3\times$). The two-stage pipeline at a similar latency, $N_1=2{,}000$ and $55.2$ ms, drops nDCG by $0.60$ and $0.66$. The two-stage pipeline with a similar accuracy, $N_1=5{,}000$, takes $115.1$ ms. S1 is constant in the query length. S2 and S3 grow with it. A query of about $400$ tokens takes $439.9$ ms and is still $29\times$ faster than full int8 MaxSim.

\subsection{Token alignment and gate weights}

\begin{figure}[tbp]
\centering
\includegraphics[width=0.90\linewidth]{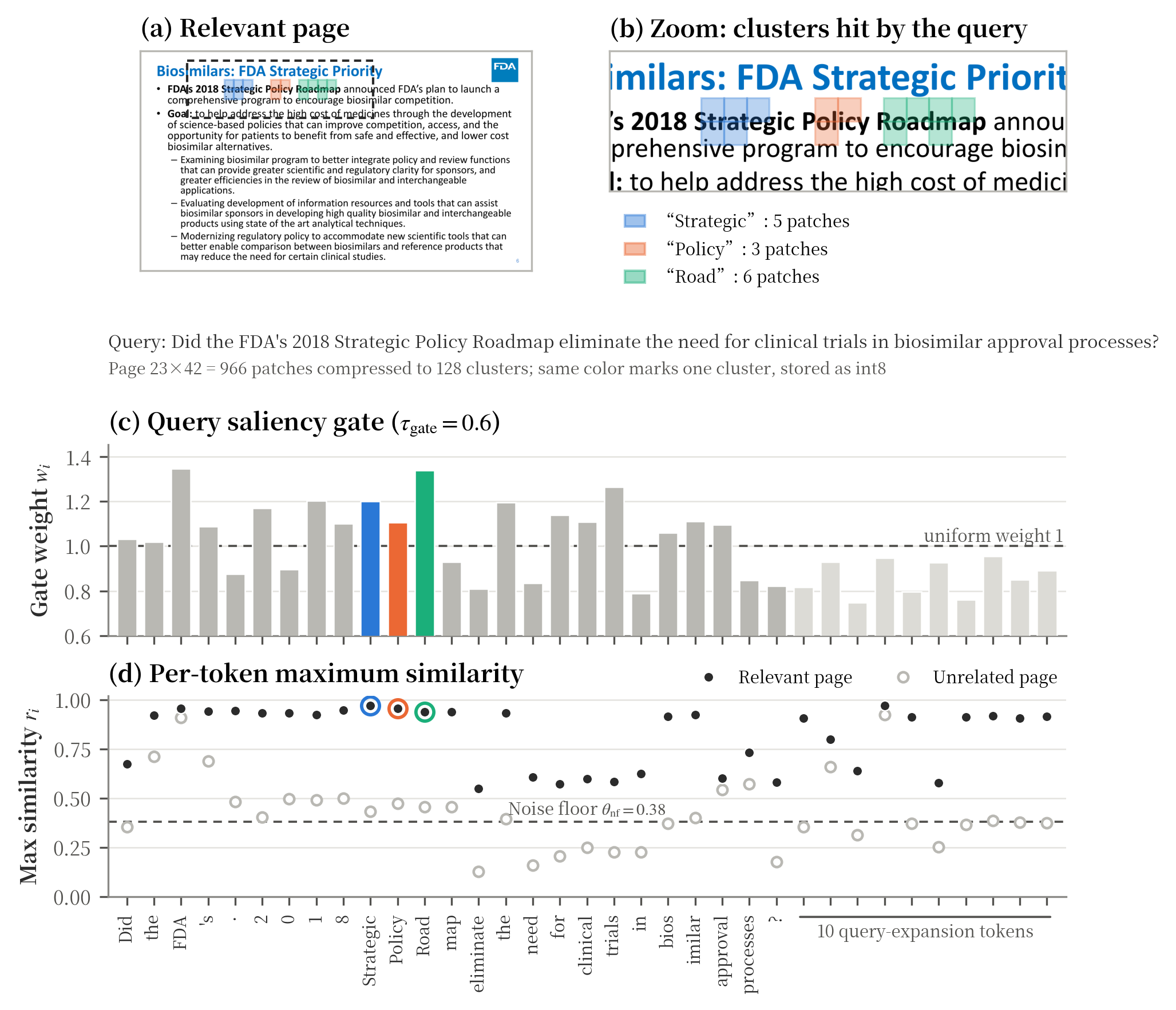}
\caption{Per-token hits and query-side gate weights. (a)(b) Member patches of the clusters hit by three query words on a medical document page from ViDoRe V3, with each cluster colored separately. (c) Per-token weights from the salience gate, with the dashed line marking the uniform weight. (d) Per-token maximum similarity on the relevant page versus an unrelated same-domain page, with the dashed line at the noise floor $\theta_{\mathrm{nf}}=0.38$.}
\label{fig:align}
\end{figure}

A single vector returns one scalar. The multi-vector index shows which part of the candidate each token hits. Figure~\ref{fig:align} uses one medical query from ViDoRe V3: ``Did the FDA's 2018 Strategic Policy Roadmap eliminate the need for clinical trials in biosimilar approval processes?'' The tokens Strategic, Policy, and Road each hit the cluster of the matching word in the bold title. Those clusters contain $3$ to $6$ patches, and the maximum similarity is above $0.94$. The gate lifts these three tokens above the uniform weight and puts the last $10$ expansion tokens below it. On the relevant page every token is above the noise floor, with mean $0.82$. On an unrelated page from the same domain the mean is $0.43$, and about half of the tokens fall below the floor. A word that appears on both pages, such as FDA, still matches. Equation~(\ref{eq:nf}) amplifies the excess above $0.38$, and a token below the floor keeps its original value.

\subsection{Evidence localization and video time localization}
\label{sec:analysis-loc}

\begin{figure}[p]
\centering
\includegraphics[width=0.90\linewidth]{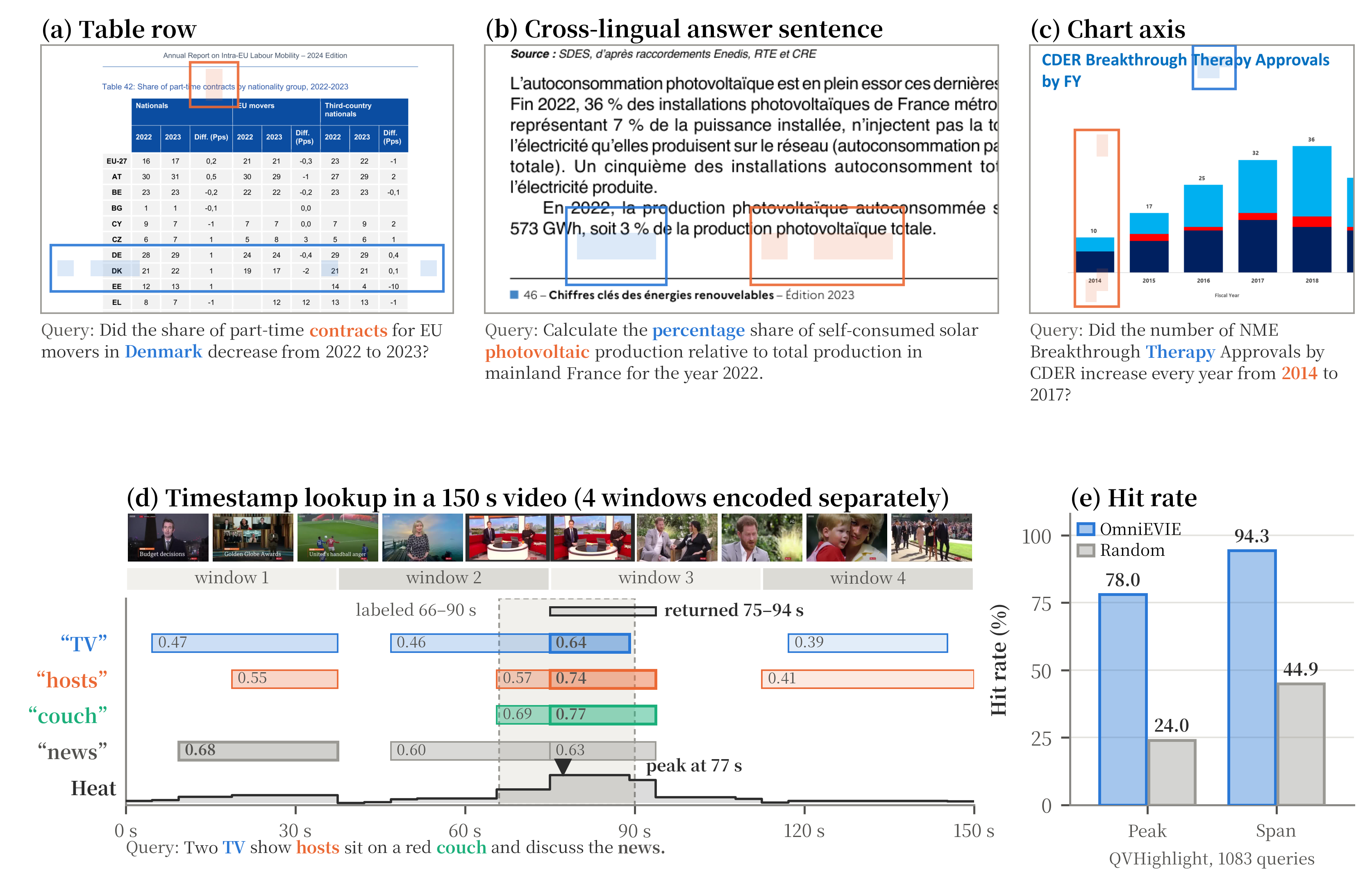}
\caption{Localized evidence. (a)(b)(c) Clusters hit on the top-ranked page: one color is one cluster, the box is its bounding rectangle, and the bold query words match the boxes. (d) Timestamp lookup in a $150$ s video: bars are each query word's best centroid in each window, shaded by the margin over the noise floor; the curve is the heat of \eqref{eq:heat}, the band is the labeled segment, and the outlined bar is the returned span. (e) Hit rate of the peak and of the returned span, against a random span of the same length.}
\label{fig:loc}
\end{figure}

The same argmax places the hit on a page region or a video time (Section~\ref{sec:loc}). In Figure~\ref{fig:loc}(a)(b)(c), three ViDoRe V3 queries hit the answer on the page ranked first by S3: Denmark hits the DK row and contracts hits the column header; the page in (b) is French, where percentage hits ``3 \%'' and photovoltaic hits ``photovolta\"{i}que totale'' in the same sentence; in (c), 2014 hits the 2014 column of the bar chart and Therapy hits the chart title.

Figure~\ref{fig:loc}(d) is a $150$ s QVHighlights video \citep{lei2021qvhighlights}, encoded in $4$ windows. The labeled segment of $66$--$90$ s crosses the window boundary at $75$ s. Couch scores $0.77$ and $0.69$ on the two sides and hosts scores $0.74$ and $0.57$, all above the noise floor, so \eqref{eq:heat} adds them; news hits another shot at $9$--$38$ s. The heat peaks at $77$ s, and the returned span of $75$--$94$ s overlaps the label and runs $4$ s past it. Over all $1{,}083$ queries, the peak falls inside the label for $78.0\%$ and the span intersects it for $94.3\%$. The span averages $27.7$ s, $18\%$ of the video, against $44.9\%$ for a random span of the same length and $24.0\%$ for a random point (Figure~\ref{fig:loc}(e)).

A global vector returns the whole video or the whole page, so a question about which second or which part of the page still sends that whole video or page to the downstream generator. An OmniEVIE hit carries its place: a bounding rectangle for a page or an image, and the returned span of \eqref{eq:heat} for a video. The downstream generator reads only that rectangle or that span; Figure~\ref{fig:loc}(a) passes the row for DK, and Figure~\ref{fig:loc}(d) passes $75$--$94$ s. The video side also needs no video parsing model. A common long-video pipeline first runs a video understanding model to transcribe each segment into timestamped captions and then retrieves over the text. OmniEVIE encodes the sampled frames directly and takes the time interval of each centroid from the frame indices, so neither indexing nor localization goes through transcription; this is a native route to video retrieval. A long video is stored as fixed windows, one record per window, and \eqref{eq:heat} returns the start and end of the matched segment on the full video timeline.

\begin{figure}[p]
\centering
\includegraphics[width=0.90\linewidth]{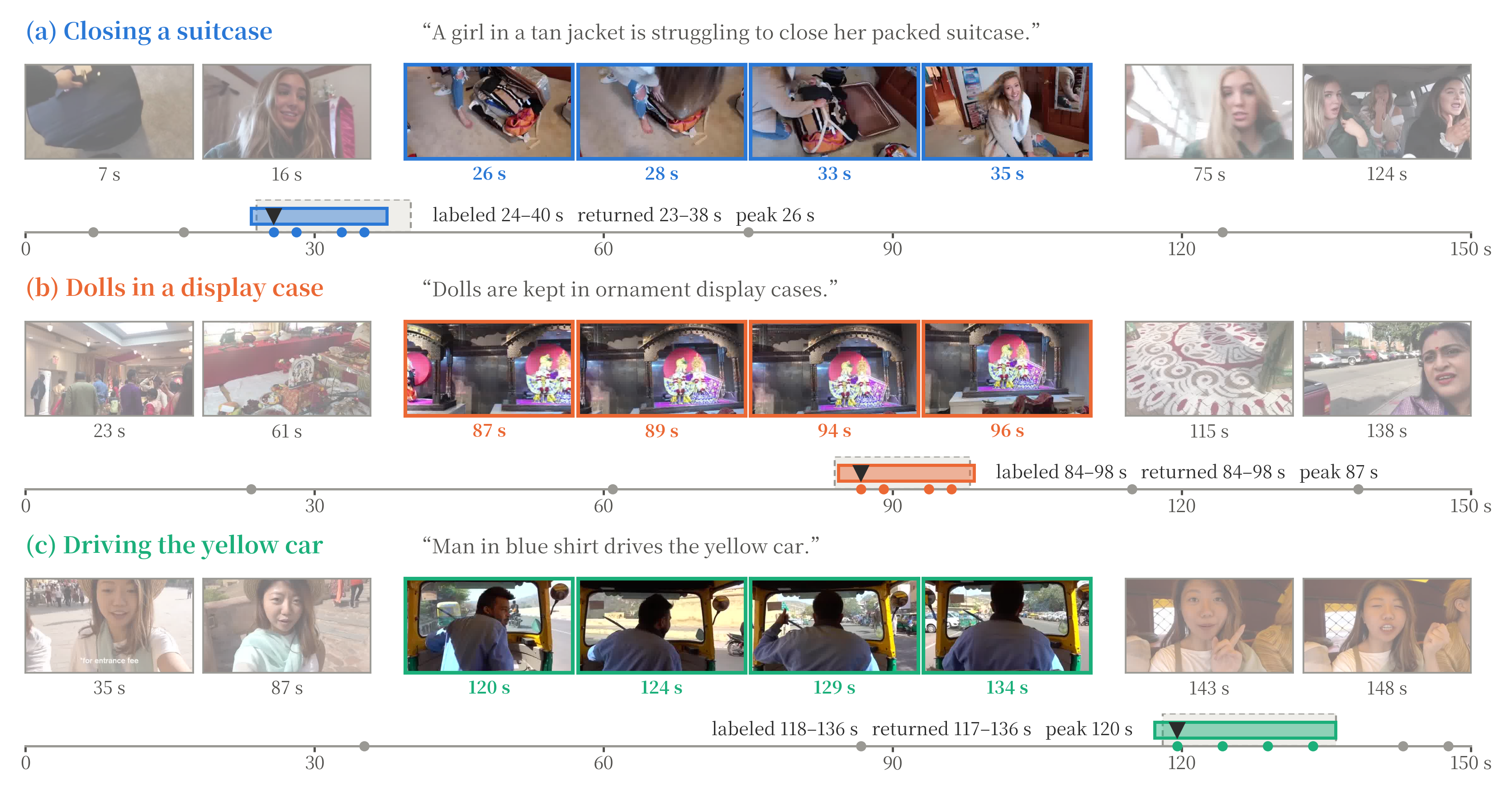}
\caption{Returned spans for three QVHighlights queries. The outlined middle frames come from the returned span and the gray side frames come from outside it; the number under a frame is its time and the dots mark these frames. The bar is the returned span, the dashed box is the labeled segment, and the triangle is the heat peak. (a) Closing a suitcase. (b) Dolls in a display case. (c) Driving the yellow car.}
\label{fig:vcases}
\end{figure}

Figure~\ref{fig:vcases} shows three QVHighlights queries, with the middle frames of each row taken from the returned span and the side frames from outside it. In (a), closing a packed suitcase, the label is $24$--$40$ s, the span is $23$--$38$ s, and the peak is at $26$ s, where she presses the clothes down into the case; outside are a close view of the bag, her facing the camera, a close-up of her face, and the two of them in a car. In (b), dolls in a display case, the label and the span are both $84$--$98$ s with the peak at $87$ s; the case fills the span, and the sides show a banquet hall, piled goods, a top-down view of a circular floor pattern, and a selfie of the woman. In (c), a man in a blue shirt driving a yellow car, the label is $118$--$136$ s, the span is $117$--$136$ s, and the peak is at $120$ s; the span shows the driver from the back seat, and the sides are the same woman's selfies.

\FloatBarrier
\subsection{Limitations}

A clip of a few seconds, sampled at a low frame rate, collapses to a handful of temporal groups under the stride of $2$, so little of the action remains. The frame rate can be chosen from the length of the clip. Reasoning retrieval, GUI, and memory are still the lowest text and agent subgroups; they need multi-step reasoning retrieval data and interface screenshots. The gate and the noise floor are computed inside S3, so their cost grows linearly with $L_q$.

\section{Conclusion}

OmniEVIE serves text, images, documents, video, and tools from one $4.5$B model and one index. MMEB-V3 is $60.06$, $0.62$ below the full $d=2048$ sequence at $60.68$ on the same 190 tasks. A record is at most $9{,}472$ bytes, about $8.8$ GiB per million records. On one million and five million records, a single thread takes $54.2$ ms and $120.2$ ms, $11.1\times$ and $25.3\times$ faster than exact search over a $4096$-dimensional vector. Search uses the compressed index. The hit names a page region or a video interval.

Training and search use the same score. Evidence grades decide which negatives are false. Candidates are clustered during training into the vectors that will be stored. Specialists trained separately are merged by parameter block, and each block receives its own coefficient. Online, the score is an inner product, a popcount, or an int8 dot product.

An agent that retrieves several times in one task benefits when each hit already names its evidence: the generator then reads that region or interval (Figure~\ref{fig:loc}). This release tests the architecture, the schedule, and the compressed index on public data: $57$ datasets and about $2$M samples. The next stage adds domain data, extends the same record to audio and to long mixed documents, and passes the localized region or interval to a generator. At larger scale the mean vector can itself be quantized or placed in a graph index.

Weights are at \url{https://huggingface.co/tencent/OmniEVIE-4B}. Inference, indexing, and the three-stage retriever are at \url{https://github.com/Tencent/OmniEVIE}. The corpora are public benchmarks and licensed document, image, and video collections. The evidence check filters sensitive material and personally identifiable information. Hyperparameters and evaluation suites are listed in Appendix~\ref{app:hparams} and Appendix~\ref{app:suites}.

\FloatBarrier
\bibliography{evie_v2}

\clearpage
\appendix

\section{Hyperparameters}
\label{app:hparams}

\begin{table}[h]
\centering
\scriptsize
\setlength{\tabcolsep}{3pt}
\caption{Training and index settings used for the released model.}
\begin{tabular}{p{0.22\linewidth}p{0.74\linewidth}}
\toprule
Group & Setting \\
\midrule
Backbone & Qwen3.5-VL with bidirectional full attention; $d_{\mathrm{hid}}=2560$; $W\in\mathbb{R}^{2048\times 2560}$; widths $\{64,128,256,512,1024,2048\}$. \\
Video & $\Delta t_{\mathrm{patch}}=2$. Training pads clips shorter than $12$ frames up to $12$, with a cap of $3{,}600$ tokens. Evaluation uses $32$ frames and $14{,}400$ tokens, and the official comparison uses $8$ frames and $3{,}600$ tokens. \\
Optimization & AdamW, weight decay $0.02$. LoRA rank $64$ or $128$, $\alpha=4r$, dropout $0.05$. Learning rate $2\times 10^{-5}$ on LoRA and $4\times 10^{-6}$ on the merger and the head, cosine decay to $0.1\times$ the peak, warmup $5\%$ of steps. Global batch $512$. One epoch per stage. \\
Loss & $\tau=0.02$, $\gamma=1$, $\beta=0.5$, $\lambda_{\mathrm{ord}}=0.10$. $\lambda_{\mathrm{kd}}$ is frozen after it is set from the early gradient-norm ratio. $w_e=0.80$. \\
Train-time $K$ & $\{64,128\}$ with probabilities $0.3$ and $0.7$. \\
Query side & $\tau_{\mathrm{gate}}=0.6$, $\theta_{\mathrm{nf}}=0.38$, $\beta_{\mathrm{nf}}=2.5$. \\
Merge & $c_{k,b}\in[0,1]$ per block. Language-layer trim keeps $\rho=0.75$; other blocks use $\rho=1$. \\
ST-HAC & Video $(w_s,w_t)=(0.07,0.03)$; images and pages $(0.10,0)$; text $(0,0.10)$. $K\le 128$, $d=64$. Time frequencies in $[\pi/4,\pi]$. \\
Quantization & Scale $1/127$, integers in $[-127,127]$. Record caps $8{,}448$ bytes (two-level) and $9{,}472$ bytes (three-stage). \\
Retrieval & $N_1=5{,}000$, $N_2=1{,}000$, $k=10$. \\
\bottomrule
\end{tabular}
\end{table}

\FloatBarrier
\section{Evaluation suites}
\label{app:suites}

\begin{table}[H]
\centering
\footnotesize
\setlength{\tabcolsep}{3pt}
\caption{Suites reported in Section~8. MMEB-V3 scores 11 audio tasks as zero and includes them in the denominator. Agent is Tool + GUI + Memory, with MCMR reported separately in Table~\ref{tab:mmebv3}.}
\begin{tabular}{p{0.28\linewidth}p{0.40\linewidth}rp{0.18\linewidth}}
\toprule
Suite & Coverage & Tasks & Metric \\
\midrule
MMEB-V3 & Image, video, visual documents, text, tool, GUI, memory, multi-hop & 190 & Hit@1; nDCG@5 on text and visual documents \\
MMEB-V2 & Image (36), video (18), visual documents (24) & 78 & Hit@1 on image and video; nDCG@5 on visual documents \\
ViDoRe V3 & 8 domains, 6 languages & 48 & Graded nDCG@10 \\
Fine-grained & FORB, Street2Shop, MovingFashion; MSR-VTT, MSVD, DiDeMo, VATEX; ActivityNet order & 8 & Top-1 / nDCG@10 / win rate \\
\bottomrule
\end{tabular}
\end{table}

Video in the fine-grained suite is encoded at $1$ frame per second, at most $32$ frames, as one native sequence.

\end{document}

%% file: math_commands.tex
\usepackage{amsmath,amsfonts,bm}

\def\eqref#1{equation~\ref{#1}}
\def\1{\bm{1}}

\DeclareMathAlphabet{\mathsfit}{\encodingdefault}{\sfdefault}{m}{sl}
\SetMathAlphabet{\mathsfit}{bold}{\encodingdefault}{\sfdefault}{bx}{n}